\documentclass[11pt]{article}

\usepackage{amsthm}
\usepackage{amsmath}
\usepackage{amssymb}
\usepackage{bm} 
\usepackage{bbm}
\usepackage{mathrsfs}
\usepackage{mathtools}
\usepackage{dsfont}
\usepackage{float}
\usepackage{bm}
\usepackage{natbib}
\setcitestyle{authoryear,open={(},close={)}}

\usepackage{graphicx}
\usepackage{adjustbox}
\usepackage{booktabs}
\usepackage{multirow}
\usepackage{caption}
\usepackage[dvipsnames]{xcolor}
\usepackage{comment}
\usepackage{nicefrac}
\usepackage{enumitem}
\usepackage{apptools}
\AtAppendix{\counterwithin{algorithm}{section}}
\AtAppendix{\counterwithin{lemma}{section}}
\AtAppendix{\counterwithin{corollary}{section}}
\AtAppendix{\counterwithin{proposition}{section}}
\AtAppendix{\counterwithin{theorem}{section}}
\AtAppendix{\counterwithin{figure}{section}}
\AtAppendix{\counterwithin{table}{section}}
\AtAppendix{\counterwithin{remark}{section}}

\usepackage{algorithm}
\usepackage{algpseudocode}
\usepackage{nicefrac}

\DeclareMathOperator*{\argmin}{arg\,min}

\usepackage{subcaption}
\numberwithin{equation}{section}
\theoremstyle{plain}

\theoremstyle{definition}

\usepackage{xspace}
\usepackage[nolist,smaller]{acronym}  
\PassOptionsToPackage{hyphens}{url}\usepackage[colorlinks=true, allcolors=blue]{hyperref}

\newcommand{\acro}[1]{\textsc{#1}\xspace }

\usepackage{xspace}
\usepackage[nolist,smaller]{acronym}  

\PassOptionsToPackage{hyphens}{url}\usepackage[colorlinks=true, allcolors=blue]{hyperref}

\newcommand{\HMM}{\acro{\smaller HMM}}%Hiden Markov Model
\newcommand{\HMMs}{\acro{\smaller HMM\footnotesize{s}}}%Hiden Markov Models
\newcommand{\HSMM}{\acro{\smaller HSMM}}%Hidden Semi Markov Model
\newcommand{\HSMMs}{\acro{\smaller HSMM\footnotesize{s}}}%Hiden semi Markov Models
\newcommand{\MCMC}{\acro{\smaller MCMC}}%Markov Chain  Monte carlo 
\newcommand{\HMC}{\acro{\smaller HMC}}%Hamiltonian Monte Carlo
\newcommand{\NUTS}{\acro{\smaller NUTS}}%No U-Turn Sampler
\newcommand{\VAR}{\acro{\smaller VAR}}%No U-Turn Sampler
\newcommand{\VARslide}{\acro{\smaller VARslide}}%No U-Turn Sampler

\newcommand{\LP}{\acro{\smaller LP}}
\newcommand{\NLP}{\acro{\smaller NLP}}
\newcommand{\LASSO}{\acro{\smaller LASSO}}
\newcommand{\TVD}{\acro{\smaller TVD}}

\newcommand{\stan}{\textit{stan}}

\def\@fnsymbol#1{\ensuremath{\ifcase#1\or *\or \dagger\or \ddagger\or
   \mathsection\or \mathparagraph\or \|\or **\or \dagger\dagger
   \or \ddagger\ddagger \else\@ctrerr\fi}}
\newcommand{\ssymbol}[1]{^{\@fnsymbol{#1}}}

\let\OLDthebibliography\thebibliography
\renewcommand\thebibliography[1]{
  \OLDthebibliography{#1}
  \setlength{\parskip}{0pt}
  \setlength{\itemsep}{3pt plus 0.3ex}
}

\def\*#1{\bm{#1}} %\def\*#1{#1}
\usepackage[left=2.3cm, right=2.3cm, top=3cm]{geometry}
\usepackage{authblk}
\title{\bf  {Bayesian Sparse Vector Autoregressive Switching Models with Application to Human Gesture Phase Segmentation }}

\author[1]{Beniamino Hadj-Amar}
\author[2]{Jack Jewson}
\author[1]{Marina Vannucci}
\affil[1]{Department of Statistics, Rice University, TX 77005-1827}
\affil[2]{Department of Economics and Business, Universitat Pompeu Fabra, Barcelona, Spain, 08005}
\affil[ ]{\textit {\textcolor{black}{beniamino.hadj-amar@rice.edu, jack.jewson@upf.edu, marina@rice.edu}}}%, 
\date{March 2024}

\begin{document}

%\doparttoc % Tell to minitoc to generate a toc for the parts
%\faketableofcontents % Run a fake tableofcontents command for the partocs

%\part{} % Start the document part
%\parttoc % Insert the document TOC

%\bibliographystyle{natbib}

\def\spacingset#1{\renewcommand{\baselinestretch}%
{#1}\small\normalsize} \spacingset{1}

\setcounter{Maxaffil}{0}
\renewcommand\Affilfont{\itshape\small}

\spacingset{1.42} % DON'T change the spacing!

\maketitle
\begin{abstract}
% In neural data applications, it is of great interest to estimate hidden states of the brain as well as temporal and spatial dependencies between locations within these hidden states. 

% Gesture analysis has been widely used for developing new methods of human-computer interaction. In gesture analysis research, it is of great interest to identify and characterize, in automated fashion, tasks related to gesture phases, which translates to  estimating hidden states from physiological measurements as well as temporal and spatial dependencies between locations within these hidden states.

We propose a sparse vector autoregressive (\VAR) hidden semi-Markov model (\HSMM)  for modeling temporal and contemporaneous (e.g. spatial) dependencies in multivariate nonstationary time series. The \HSMM's generic state distribution is embedded in a special transition matrix structure, facilitating efficient likelihood evaluations and arbitrary approximation accuracy. To promote sparsity of the \VAR coefficients, we deploy an $l_1$-ball projection prior, which combines differentiability with a positive probability of obtaining exact zeros, achieving variable selection within each switching state. This also facilitates posterior estimation via Hamiltonian Monte Carlo (\HMC).   We further place non-local priors on the parameters of the \HSMM dwell distribution improving the ability of Bayesian model selection to distinguish whether the data is better supported by the simpler hidden Markov model (\HMM), or the more flexible \HSMM. Our proposed methodology is illustrated via an application to human gesture phase segmentation based on sensor data, where we successfully identify and characterize the periods of rest and active gesturing, as well as the dynamical patterns involved in the gesture movements associated with each of these states.
\end{abstract}

\noindent % 
%{\it Keywords:}  Markov Switching Process; Hamiltonian Monte Carlo;  Bayes Factor;  Telemetric Activity Data; Circadian Rhythm.
\spacingset{1.45} % DON'T change the spacing!

\section{Introduction}
Vector autoregressive (\VAR) models offer a principled methodology for detecting the dynamics and connections between multiple time-series \citep{sims1980macroeconomics, lutkepohl2005new}, wherein the observed data is assumed to be a noisy linear combination of some finite set of past observations. \VAR models have proven to be especially useful in applications to complex data, for example in the analysis of the dynamical behaviours of economic and financial time series \citep{watson1994vector, ahelegbey2016bayesian, kalli2018bayesian}, and of high-dimensional signals arising from different areas of the brain, such as fMRI \citep{goebel2003investigating} and EEG data \citep{prado2006multivariate, kammerdiner2010analysis}. Nonstationarity is commonly observed in many physiological time series, as a result of external perturbations or due to an individual performing distinct tasks or experiences \citep{ombao2018statistical, hadj2021identifying}. 
In the application of this paper, we consider multivariate time series data that arise from a study on human gesture phase segmentation based on sensor data. Phase segmentation aims at identifying and characterising the principal units, the underlying hidden states of interest to practitioners, and the dynamical patterns involved in the gesture movements associated with each of these states %during each state an individual's gesture sequence \benni{What do we mean by gesture activity, I don't think this means anything on its own, ``hand gestures when telling a story''?} 
\citep{moni2009hmm, wagner2014gesture}. Here, we analyse scalar velocity and acceleration recordings  collected at discrete time intervals on the left hand, right hand, left wrist, and right wrist of an individual who is asked to read two distinct comic strips and to tell the stories in front of a sensor \citep{madeo2013gesture}.  
As a segmentation exercise, we aim to model the data to identify, in an automated fashion, the periods of rest and active gesturing. %\benni{for AoAS do we need to say why this is useful, I guess we want to automate what was currently done by hand - why might this be important? Does it say in the original paper?} 
%\jack{Yes, in the paper they say: "Moreover, gesture analysis may be also used for developing tools aiming at helping linguists to analyze the interaction between speech, gestures, and discourse. These tools could automate some laborious or time-consuming tasks within their process of gesture analysis. One of these tasks is the segmentation of gesture phases." Also they say that they not found many studies focusing on the automation of these tasks.}
Automating such a classifications circumvents a human being from conducting such a laborious, time consuming and ultimately expensive task and allowing linguists greater resources to investigate the interaction between speech, gestures and discourse.

\textcolor{black}{There are two main tasks when analysing this data, which motivated the methodological development of our proposed model. The first is to \textit{classify} whether a particular gesture phase corresponds to a resting or active state. While \cite{sarkar2023human} considered several supervised machine learning methods for this task, \cite{wagner2014gesture} reported ambiguity in the problem of analysing gestures and serious disagreement, even among experts, in determining when a gesture unit starts and ends. It is, therefore, of interest whether unsupervised, hidden state models can replicate supervised classification labels and help address these differences of opinion in a data driven fashion. A second main task of the analysis is to \textit{investigate the dynamic associations} of acceleration and velocity measurements collected from left hand (LF), right hand (RH), left wrist (LW), and right wrist (RW) and how they differ between resting and active states. 
%These applied considerations required us to combine several recent methodological contributions in a nocel manner
We combine several recent individual works to achieve these goals. In particular, our unified approach considers: (i) hidden state models with \VAR emissions and general dwell distributions; (ii) non-local priors for improved selection between nested dwell distributions; (iii) the $l_1$-ball projection prior for model selection and estimation of the \VAR parameters.
}

% , and ... \benni{can we have one more thing about why this is useful, do they use this for speech therapy, people with special educational needs, something like that to make it seem important}
% However, until now there has been an absence of methods facilitating such automation. \benni{add reference here}

\VAR Hidden Markov Models (\VAR \HMM), also called \VAR Markov switching processes \citep{hamilton1989new, fox2014joint, samdin2016unified}, have proven useful in modeling nonstationary streams of data. Formally, a \VAR-\HMM is a stochastic process which is based on an unobserved (hidden) state sequence that takes discrete values and whose transition probabilities follow a Markovian structure.  Conditioned on the state sequence, the observations are assumed to be generated from a state-specific \VAR process. 
%\jack{. For the first time we consider the generalised \HSMM model with \VAR emissions and make this feasible using the likelihood approximation of \cite{langrock2011hidden}. }
Although computationally convenient, assuming that the hidden states are Markovian limits the flexibility of \HMMs. In particular, the dwell duration in any state, namely the number of consecutive time points that the time series spends in that state, implicitly follows a geometric distribution, an assumption that has been shown to be unrealistic in many data applications  \citep{zen2007hidden, pimentel2015heart, hadj2022bayesian}. Hidden semi-Markov Models (\HSMMs) provide a more general framework by introducing an explicit, state specific, form for the dwell duration \citep{guedon2003estimating}. 
\textcolor{black}{We consider for the first time \HSMM models with \VAR emission distributions and facilitate Bayesian inference \citep{hadj2022bayesian} via the likelihood approximation introduced by \citet{langrock2011hidden}, in which a special structure of the transition matrix is used to model the state duration distributions.}
%The state then stays unchanged until the duration terminates, at which point there is a Markov transition to a new regime. While their flexibility is appealing, Bayesian inference for \HSMMs had long been plagued by the substantially increased computational cost of computing its likelihood: the message-passing procedure for  \HSMMs requires $O(T^2K +T K^2)$ basic computations for a time series of length $T$ and number of states $K$, whereas the corresponding forward-backward algorithm for an \HMM has a computational complexity of $O(TK)$. 
%Recently, to facilitate Bayesian inference for \HSMMs, \citet{hadj2022bayesian} deployed the likelihood approximation introduced by \citet{langrock2011hidden}, in which a special structure of the transition matrix is used to model the state duration distributions. 
\textcolor{black}{This allows us to select, in a data driven manner, which dwell distribution is best supported by the data and therefore leads to improved state classification.}

\textcolor{black}{To expedite this selection, we place non-local priors \citep{johnson2012bayesian} on the parameters of the \HSMM dwell distribution, improving the ability of Bayesian model selection to distinguish whether the data is better supported by the simpler \HMM, or a more flexible \HSMM. While non-local priors have been previously deployed for linear regression \citep{rossell2017nonlocal} and Gaussian mixture modelling \citep{fuquene2019choosing} we extend this to more general nested model selection problems.}
%Lastly, we place non-local priors \citep{johnson2012bayesian} on the parameters of the \HSMM dwell distribution, improving the ability of Bayesian model selection to distinguish whether the data is better supported by the simpler \HMM, or a more flexible \HSMM. 

Capturing the complex multivariate dependencies in  human gesture data applications requires modelling not only the serial dependence within each univariate series, but also the interdependence across distinct physiological measurements, both within and across time points. 
To achieve this, we consider \VAR \HSMM models that allow temporal and contemporaneous (e.g. spatial) dependencies in the non-stationary time series \citep{paci2020structural,hadj2022bayesian}.  
%A switching \VAR process, in particular, is appropriate to characterize such time-varying dependence configurations, where the autoregressive parameters are fundamental in understanding the temporal dynamics. 
\textcolor{black}{\VAR models are generally overparametrised, allowing for dependence between all dimensions of the multivariate time series where the true dependence structure is often much sparser. 
%Understanding the temporal dynamics of the data therefore requires a sparse \VAR representation. 
%Estimating a certain \VAR coefficient to be zero means that a time series at a previous lag was not predictive of the other, i.e. that one does not Granger cause the other \citep{granger1963economic}. 
Shrinkage priors, such as the Minnesota prior \citep{doan1984forecasting} and numerous variations \citep{kadiyala1997numerical, lutkepohl2005new}, the doubly adaptive elastic-net \LASSO \citep{gefang2014bayesian}, and the Bayesian nonparametric \LASSO \citep{billio2019bayesian} have been used for \VAR coefficients, but, often arbitrary, post-hoc thresholding is required to estimate the zeros.}
%In econometric applications, the Minnesota prior by \citet{doan1984forecasting}  was introduced  as a shrinkage prior for autoregressive parameters and numerous variations have been presented since \citep{kadiyala1997numerical, lutkepohl2005new}. \citet{gefang2014bayesian} developed a Bayesian doubly adaptive elastic-net \LASSO  approach for \VAR shrinkage and \citet{billio2019bayesian} presented a Bayesian nonparametric \LASSO prior for high dimensional \VAR models. 
%Model selection methods, on the other hand, have the added advantage of shrinking coefficients exactly to 0 and revealing interpretable structures. 
% Spike-and-slab variable selection priors \citep{george2008bayesian} could be specified to identify the non-zero \VAR coefficients in situations where there is an interest in learning a graphical network among the measured high-dimensional signals, such as a connectivity network among brain regions in fMRI studies \citep{Chiang2017, Kook2021}.
\textcolor{black}{A common framework to introduce exact zeros into a Bayesian analysis is to use spike-and-slab priors \citep{george1993variable,chiang2017bayesian}.}
\textcolor{black}{Here, we deploy, for the first time, the novel $l_1$-ball projection prior \citep{xu2020bayesian}, a prior with mass at exact zeros, on the coefficients of the \VAR matrices.
Such a prior introduces unconstrained latent variables and transforms them onto the space of \VAR parameters in a way that provides positive probability that any element is exactly zero. Importantly,  
the transformation is almost surely continuous and differentiable, allowing us to estimate edge inclusion probabilities 
%and to test for Granger causality, 
while remaining compatible with efficient posterior sampling using algorithms such as Hamiltonian Monte Carlo (\HMC, \citealt{duane1987hybrid}) and off-the-shelf probabilistic programming languages such as \stan{} \citep{carpenter2016stan}.}
%The resulting Markov Chains will contain a mixture of elements that are exactly zero as well as non-zeros facilitating straightforward estimation of posterior inclusion probabilities without any thresholding or post-processing. Lastly, 
%We embed the \HSMMss generic state distribution in a special transition matrix structure, facilitating efficient likelihood evaluations and arbitrary approximation accuracy, as in \citet{hadj2022bayesian}. 
%Lastly, we place non-local priors \citep{johnson2012bayesian} on the parameters of the \HSMM dwell distribution, improving the ability of Bayesian model selection to distinguish whether the data is better supported by the simpler \HMM, or a more flexible \HSMM. 
\textcolor{black}{Lastly, we place standard priors on the state specific covariance matrices that capture contemporaneous spatial dependencies, taking advantage of \stan{} implementations that use the Cholesky decomposition. }

We show on simulated data how our model is able to detect sparse \VAR structure across hidden switching regimes as well as estimate contemporaneous dependencies and conduct model selection between the simpler \HMM and the more flexible \HSMM model. \textcolor{black}{Applied to the gesture phase data, 
%we show that our unsupervised model is well able to classify states as active and resting, improving upon the analogous \HMM model, as well as estimating interpretable \VAR structures with differing levels of sparsity across the states. 
unsupervised learning appears to perform well for resting and active state classification, showing potential to address some of the `disagreements between experts' and dwell times in these states appear better modelled by a negative binomial distribution than a geometric distribution, achieving better model fit and state classification.}
Further, our model also 
estimated interpretable \VAR structures with differing levels of sparsity across the states and
%{\color{red}We show on simulated data ...... In the application, ...}
%develop a probabilistic modelling framework that is not only able to describe and classify events in the observed time series, but that it is also capable of 
successfully characterises and predicts a new video, corresponding to a different, unseen, story.

The rest of the paper is organised as follows: Section \ref{seq:model} introduces our \VAR HSMM model, the $l_1$-ball prior utilized to impose sparsity on the temporal connections and our strategy for posterior inference. Section \ref{Sec:Simulations} contains our simulation study, including comparisons of the selection performance of local and non-local priors.  Lastly, Section \ref{Sec:GesturePhase} presents our application to the gesture phase data and Section \ref{sec:summary_discussion} concludes the paper with a discussion. The \stan{} files (and  \texttt{R} utilities) that were used to implement  our analysis are available at 
\url{https://github.com/Beniamino92/sparseVARHSMM}.  %The probabilistic programming framework associated with \stan{} makes it easy for practitioners to consider further dwell/emission distributions to the ones considered in this paper. Users need only change the corresponding function in our \stan{} files.
We use the probabilistic programming framework associated with \stan{}, which facilitates extensions of the code to similar applications. 

%describes our Bayesian inference procedure: with Section \ref{Sec:VAR_l1_prior} explaining  of our \VAR and how we elicit its parameters, Section \ref{Sec:CovarianceMatrix} defining our prior on the full covariance matrix of the \VAR observations and Section \ref{Sec:dwell_distributions} explaining our priors on the parameters of the dwell distribution, including our non-local priors for the additional dwell parameters introduced by the \HSMM. Section \ref{Sec:Simulations} presents some simulation studies demonstrating that...
%\benni{..I still need to finish this, when the paper is finish. it is a mini intro of the various secctions....} \benni{add the stan code and R utilies can be found on github blabla}

\section{VAR Approximate HSMMs}{\label{seq:model}}
We describe our proposed vector autoregressive (\VAR) hidden semi-Markov model (\HSMM)  for modelling temporal and contemporaneous (e.g. spatial) dependencies in high-dimensional nonstationary time series. 
%The HSMM's generic state distribution is embedded in a special transition matrix structure, facilitating efficient likelihood evaluations and arbitrary approximation accuracy. We deploy an $l_1$-ball projection prior, which combines differentiability with a positive probability of obtaining exact zeros, to conduct variable selection within each different switching regime and facilitate posterior estimation via HMC (\textit{stan}).\jack{Does this paragraph not just repeat the intro?}

\subsection{Vector Autoregressive Emissions}
\label{seq:VAR_emisison}

Let  $\bm{y} = \big\{ \bm{y_t} \big\}_{\, t= 1}^{T}$ be $D$ dimensional observed time series, with $\bm{y}_t = (y_{t1}, \dots, y_{tD}) \in \mathbb{R}^{D}$, and let $\bm{z} = (z_1, \dots, z_T)$ indicate the hidden states associated with each time point, with $z_t \in \{1, \dots, K\}$ and $K \in\mathbb{Z}_{+}$ the number of latent states. Conditional on the value of the hidden state sequence $z_t$,  the data  $\bm{y}_t$ are modelled through a state-specific \VAR process of order $P \in\mathbb{Z}_{+}$, which assumes $\bm{y}_t$ to be a linear combination of past observations $\bm{y}_{t-1:t-P}$, across all dimensions, plus a Gaussian noise term, as
 %,  the latter being an integer denoting the number of autoregressive terms considered in the model.  The generative process of the state-specific emissions can be expressed as
    \begin{equation}
    \bm{y}_{t}
\, | \,  \textcolor{black}{z_{\,t} = j} \, \sim \, \mathcal{N}_D \, \Big( \bm{\alpha}^{\,j} + \sum_{p=1}^{P} \bm{\Theta}_p^{\,j} \, \bm{y}_{t-p}, \, \bm{\Sigma}^{\,j} \, \Big), \qquad t = P+1, \dots, T,\label{Equ:VAR_emission}
    \end{equation}
where for each regime $j=1, \dots, K$, $\bm{\alpha}^{\,j} \in \mathbb{R}^{D}$ is a vector of state-dependent intercepts, that allow for the possibility of a non-zero mean $\mathbb{E}[\bm{y}_t | \bm{y}_{t-1:t-P}]$, and $\bm{\Theta}_{\,p}^{\,j} = \{\theta^{\,j}_{pil}\}_{i,l=1}^{D}  \in \mathbb{R}^{D\times D}$ is the  matrix of autoregressive coefficients at lag $p=1, \dots, P$. 

Unlike common approaches that assume independent noise across dimensions, for computational convenience,
%in a pseudo likelihood approach \citep{ghosh2021strong}, 
\eqref{Equ:VAR_emission} allows for contemporaneous dependencies as well as temporal dependencies via a state specific positive definite covariance matrix $\bm{\Sigma}^{\,j} = \{ \sigma^{\,j}_{il}\}$. 
%\subsection{Covariance matrices}{\label{Sec:CovarianceMatrix}}
%\jack{Notation reminder: matrix bold upper, element not bold lower}
We decompose $\bm{\Sigma}^{\,j}$ into a vector of scales $\bm{\tau}^{\,j}$ and a correlation matrix $\bm{\Omega}^{\,j} = \{\omega_{il}^{\, j}\}$, where $\tau_i^{\,j} = \sqrt{\sigma_{ii}^{\,j}}$ and $\omega^{\,j}_{il} = \frac{\sigma^{\,j}_{il} }{\tau_i\tau_l}$ \citep{gelman2006data}. 
Given standardised data, we set $\tau^{\,j}_i\sim \text{Cauchy}(0.5, 0.5)\mathbb{I}(\tau^{\,j}_i > 0)$, providing high prior probability of relatively small scales but with heavy tails to ensure that the prior does not overly regularise a regime with greater variance. Then, we specify the prior for the correlation matrices as %$\Omega^{\,j}$ are assumed a priori 
\begin{align}
    \bm{\Omega}^{\,j} \sim \text{LKJ}(\bm{\Omega}^{\,j}; \xi) \propto \text{det}(\bm{\Omega}^{\,j})^{\xi - 1}, \nonumber
\end{align} 
where LKJ denotes the prior proposed by \citep{lewandowski2009generating}, namely a family of probability distribution for positive definite correlation matrices (or equivalently for their Cholesky factors).
Setting $\xi = 1$ provides a uniform distribution over correlation matrices. Note this is different from a uniform distribution over elements of the correlation matrix, as $\bm{\Omega}^{\, j}$ is constrained to be symmetric and positive definite. \textcolor{black}{Such a choice of prior allows us to input no prior information about the strength of contemporaneous dependencies, while ensuring that the covariance matrix is positive definite}. This prior implemented in \stan{} via the Cholesky decomposition \citep{stan2022stan}.

%benni{Jack, I think you should mention something about that we model the lower triangular part of the cholesky decomposition. 
%\url{https://mc-stan.org/docs/2_18/stan-users-guide/multivariate-hierarchical-priors-section.html}}

\subsection{Introducing Sparsity in the \VAR Coefficients via the $l_1$-ball Projection Prior}{\label{Sec:l1_ball}}

%\jack{Need to motivate beyond overparametrisation (as this is solved wih shrinkage), beyod this need to motivate as Grager Causality - DISCUSSION OF OTHE RMETHODS}
%\jack{Add in Benni's tetx here - the OLS stuff.}

\textcolor{black}{Shrinkage priors are commonly used \citep{doan1984forecasting,gefang2014bayesian,billio2019bayesian} to deal with over parametrisation in \VAR models.}
%However, an ordinary switching \VAR model is often overparameterized since the number of autoregressive coefficients, in each state, grows quadratically with the dimension $D$ of the multivariate time series.
%A common issue with \VAR models is overparametrisation. 
The number of distinct elements of  $\bm{\Theta} = \{\bm{\Theta}^{\,j} 
\}_{j=1}^K$, with $\bm{\Theta}^{\,j} := \{ \bm{\Theta}^{\,j}_p \}_{p=1}^{P}$, for $j=1, \dots, K$, is $KPD^{2}$, which grows quickly even for moderate dimensional data. 
%moderate and high dimensional data, the number of possible temporal connections in each state, which grows quadratically with the dimension $D$ of the multivariate time series, is often much larger than the true number of connections \citep{davis2016sparse, paci2020structural}.
%Thus, shrinkage priors are commonly used \citep{doan1984forecasting,gefang2014bayesian,billio2019bayesian}. 
% However, in many applications, the presence or absence of temporal connections among the $D$ dimensions can be of scientific interest, and it is therefore appropriate to induce sparsity by setting connections not supported by the data to zero, a task that can be achieved, for example, via variable selection {\it spike-and-slab} priors, see \cite{BVS_book} for a comprehensive treatment of these priors. For \VAR models, \cite{Chiang2017} and \cite{Kook2021} used discrete spike-and-slab priors on the model coefficients to learn a connectivity network among brain regions in fMRI studies.  However, there are computational issues with so doing, often requiring bespoke algorithms or approximation to facilitate inference.
Here, we deploy the recently proposed $l_1$-ball projection  prior of \citet{xu2020bayesian}, which augments the sample space with strictly continuous latent parameters and applies an almost surely continuous and differentiable transformation that results in model parameters with a mixture-type prior including positive probability of exact 0's and continuous density away from 0.
%{\color{red}which the authors argue captures the spike-and-slab prior as a special case.} However, unlike discrete spike-and-slab priors, 
Continuity and differentiablity of the $l_1$-ball projection  prior allows for efficient posterior sampling using algorithms such as Hamiltonian Monte Carlo (\HMC, \citealt{duane1987hybrid}). \textcolor{black}{\cite{xu2020bayesian} note that the classic spike-and-slab prior can be viewed as a special case of the $l_1$-ball prior and  demonstrate empirically that the $l_1$-ball prior leads to faster mixing performance than discrete spike-and-slab.}

\begin{figure}[htbp]
\centering
\includegraphics[width=0.5\linewidth]{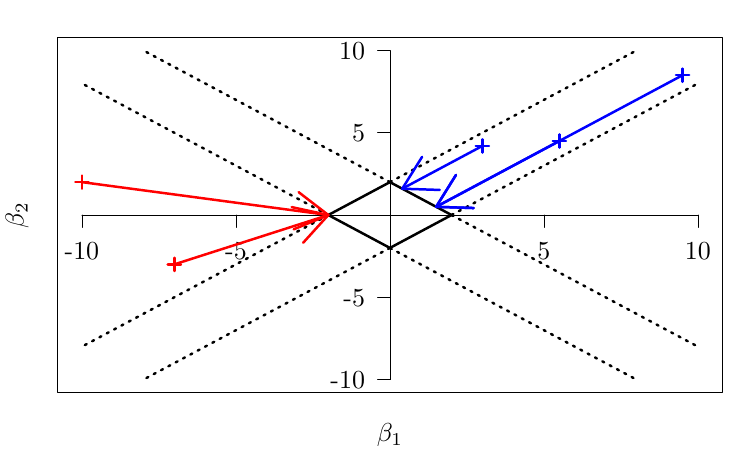}
\caption{$l_1$-ball projection in $\mathbb{R}^2 $  mapping $\bm{\beta}^{\,j}$ to $\bm{\theta}^{\,j}$ with radius $r^{j}=2$. Any $\bm{\beta}^{\,j}$'s in the dotted `V's away from the axis has one of their coordinates mapped to 0 in $\bm{\theta}^{\,j}$, while any $\bm{\beta}^{\,j}$'s in the diagonal columns going away from the origin do not have either coordinate mapped to 0.}
   \label{fig:l1_ball}
\end{figure}

%% Benni's first go %%
\textcolor{black}{
Before formally defining the $l_1$-ball projection prior we provide an intuitive illustration. 
%In order to  define the $l_1$-ball projection prior 
We introduce unconstrained latent variables, $\bm{\beta}^{\,j}$, and transform them onto the space of \VAR parameters, $\bm{\theta}^{\,j}$, in a way that provides positive probability that any element is exactly zero. The degree of sparsity in the transformation is controlled by a positive scalar-valued random variable called the radius $r^{j}$. 
%To provide an intuition behind the prior, 
Figure \ref{fig:l1_ball} displays the $l_1$-ball projection in $\mathbb{R}^2$ with $r^{j}=2$. Any $\bm{\beta}^{\,j}$ values in the dotted `V's away from the axis have one of their coordinates mapped to 0 in $\bm{\theta}^{\,j}$, e.g. the two red lines at $\bm{\beta}^{\,j} = (-10, 2)$ and $\bm{\beta}^{\,j} = (-7, -3)$ which are both mapped to $\bm{\theta}^{\,j} = (-r^{\,j}, 0)$. Alternatively, for $\bm{\beta}^{\,j}$ values in the diagonal columns going away from the origin, neither coordinate is mapped to 0 in $\bm{\theta}^{\,j}$, e.g. the blue point $\bm{\beta}^{\,j} = (3, 4.2)$ is mapped to $\bm{\theta}^{\,j} = (0.4, 1.6)$, and both $\bm{\beta}^{\,j} = (5.5, 4.5)$ and $\bm{\beta}^{\,j} = (9.5, 8.5)$ are mapped to $\bm{\theta}^{\,j} = (1.5, 0.5)$.
}

%\begin{algorithm}
%\caption{$l_1$-ball projection. $\, \, $ \textit{\textbf{Input:}}
%$\bm{\beta}_p^{\,j} \in \mathbb{R}^{D^2}$ and $r^{\,j} \in \mathbb{R}_{+}$. $\, \, $ \textit{\textbf{Output:}}
%$\bm{\Theta}_p^{\,j} \in \mathbb{R}^{D \times D} \cup \{0\}$ }\label{alg:l1_balla}
%\begin{algorithmic}
%\If{$||\bm{\beta}_p^{\,j}||_1 \leq r^{\,j}$}
%    \State $\bm{\Theta}_p^{\,j} \gets vec^{-1}(\bm{\beta}_p^{\,j})$
%\Else    
%\State Sort $\bm{\beta}_p^{\,j}$ so that $|\beta^{\,j}_{p\,(1)}| \geq \cdots \geq |\beta^{\, j}_{p\,(D^2)}|$ 
%\State $\phi_l\gets \left(\sum_{n=1}^l |\beta^{\,j}_{p\,(n)}| - r^{\,j}\right)_{+},$ $\, \qquad \quad \text{for } \, l = 1, \dots, D^2. $ 
%\State $m \gets \max\left\{n: |\beta^{\,j}_{p\,(n)}| > \frac{\phi_n}{n}\right\}$
%\State $\tilde{\phi} \gets \frac{\phi_m}{m}$
%\State $\theta^{\,j}_{p i} \gets~ \textup{sign}(\beta^{\,j}_{pi})\, \max\left(|\beta^{\,j}_{pi}| - \tilde{\phi}, 0 \right),$ $\, \, \text{for } \, l = 1, \dots, D^2. $ 
%\State $\bm{\Theta}_p^{\,j} \gets vec^{-1}(\bm{\theta}_p^{\,j})$
%\EndIf
%\end{algorithmic}
%\end{algorithm}

\textcolor{black}{Formally}, define $\bm{\beta}^{\,j}_{p} =  (\beta^{\,j}_{p1}, \dots, \beta^{\,j}_{pD^2}) = vec(\bm{B}_p^{\,j})$, with each element of matrix $\bm{B}^{\,j}_{p} = \{ \beta^{\,j}_{pil} \}_{i,l=1}^{D} \in \mathbb{R}^{D\times D}$ corresponding to the latent parameter associated with each autoregressive coefficient. Similarly, we define $\bm{\theta}^{\,j}_p =   (\theta^{\,j}_{p1}, \dots, \theta^{\,j}_{pD^2}) = vec(\bm{\Theta}_p^{\,j})$. Here, the $vec(\cdot)$ operator denotes the vectorization of a matrix into a column vector, i.e. $vec: \mathbb{R}^{D \times D} \rightarrow \mathbb{R}^{D^2}$. Further, we construct  the vector of state-specific latent parameters $\bm{\beta}^{\,j} = [\bm{\beta}_1^{\, j}, \dots, \bm{\beta}_P^{\, j}]'$ and autoregressive parameters $\bm{\theta}^{\,j} = [\bm{\theta}_1^{\, j}, \dots, \bm{\theta}_P^{\, j}]'$, by concatenating the corresponding vector of lag-specific elements.  Then, the $l_1$-ball projection can be specified as 
         \begin{equation}
   \{ \beta^{\,j}_{pil} \}_{i,l=1}^{D} \overset{\mathrm{iid}}{\sim} \text{DExp}(0, \sigma_\beta^{\,j}), \quad r^{\,j} \overset{\mathrm{iid}}{\sim} \text{Expo}(a_r^{\,j})
   \label{eq:l1_ball_1}
    \end{equation}
        \begin{equation}
        \bm{\theta}^{\,j} := \argmin_{||\bm{x}||_1 \, \leq \, r^{\,j}} ||\bm{\beta}^{\,j} - \bm{x}||_2^2, \qquad \bm{\Theta}^{\,j}_{p} := vec^{-1} \big(\bm{\theta}^{\,j}_p \big), \quad \text{for} \, \, p = 1, \dots, P, 
         \label{eq:l1_ball_2}
    \end{equation}
    %\begin{equation*}
    %    \bm{\Theta}^{\,j}_{p} := vec^{-1} \big(\bm{\theta}^{\,j}_p \big), \quad \text{for} \, \, p = 1, \dots, P,  
    %     \label{eq:l1_ball_3}
    %\end{equation*}
%     \begin{equation}
%   \{ \beta^{\,j}_{pil} \}_{i,l=1}^{D} \overset{\mathrm{iid}}{\sim} \text{DExp}(0, \sigma_\beta^{\,j}), \quad r^{\,j} \overset{\mathrm{iid}}{\sim} \text{Expo}(\alpha_r^{\,j})
%   \label{eq:l1_ball_1}
%     \end{equation}
%         \begin{equation}
%         \bm{\Theta}^{\,j}_{p} := vec^{-1} \Big( \argmin_{||\bm{x}||_1 \, \leq \, r^{\,j}} ||\bm{\beta}^{\,j}_{\,p} - \bm{x}||_2^2 \Big), 
%          \label{eq:l1_ball_2}
%     \end{equation}
for $j=1, \dots, K,$ where $||\cdot||_1$ and $||\cdot||_2$ denote the $l_1$ and $l_2$ norms, respectively, and where the state-specific radius is $r^{\,j} > 0$. %, 
%\textcolor{red}{is a positive scalar-valued random variable controlling the degree of sparsity of the autoregressive coefficients}. 
The operator $vec(\cdot)^{-1}$ represents the inverse of the vectorization operator, i.e. $vec^{-1}: \mathbb{R}^{D^2} \rightarrow \mathbb{R}^{D \times D}$ such that  $vec\,\big(vec^{-1}(\bm{\Theta}^{\,j}_{p})\big) = \bm{\Theta}^{\,j}_{p}$, $\text{Expo}(a)$ represents an exponential distribution with rate $a$, and $\text{DExp}(\mu_\beta, \sigma_\beta)$ represents a double-exponential (often called a Laplace distribution) with mean $\mu_\beta$ and scale parameter $\sigma_\beta$. %For notational convenience, define the function representing the solution to Eq. \eqref{eq:l1_ball_2} as $l_1: \mathbb{R}^{D\times D}\times \mathbb{R}_{+}\mapsto \mathbb{R}^{D\times D}$. \jack{Think about if we really need this!}
%The straightforward algorithm provided by \cite{xu2020bayesian} to solve the minimisation in \eqref{eq:l1_ball_2} in given in by Algorithm \ref{Alg:l1_ball}.

We employ the $l_1$-ball projection prior construction \eqref{eq:l1_ball_1}-\eqref{eq:l1_ball_2} independently in  each regime: a  double exponential distribution is assumed on random variables $\beta^{\,j}_{pil}$; 
%(we chose a double exponential distribution as it facilitates prior elicitation (Section \ref{Sec:l1_prior_elicitation}) but other continuous distributions are equally valid);  
the $l_1$-ball projection with radius $r^{\,j}$ is applied to map $\bm{\beta}^{\,j}\mapsto \bm{\theta}^{\,j}$ 
such that if $||\bm{\beta}^{\,j}||_1 \geq r^{\,j}$ then $||\bm{\theta}^{\,j}||_1 = r^{\,j}$ with some of the  $\theta^{\,j}_{pil} = 0$; 
%by minimizing the loss function $ ||\bm{\beta}^{\,j}_{\,p} - \bm{x}||_2^2$ on $\bm{x} \in \mathbb{R}^{D^2}$ such that $||\bm{x}||_1 \, \leq \, r^{\,j}$; 
the matrix of autoregressive coefficients  $\bm{\Theta}^{\,j}_{p}$ is reconstructed using the $vec^{-1}$ operator. 
The loss function in  Eq. \eqref{eq:l1_ball_2} is strictly convex, namely for every $\bm{\beta}^{\,j}$, there is only one optimal solution $\bm{\theta}^{\,j}$. The projection can be solved using a variant of the procedure by \citet{duchi2008efficient} and \citet{xu2020bayesian}, which is summarized in Algorithm \ref{alg:l1_balla} in the Supplementary Material.  This mapping function is almost surely continuous and differentiable, % hence allowing for compatibility with \HMC and \textit{stan}. Additionally, the $l_1$-ball  projection  yields a volume preserving transformation \citep{xu2020bayesian}, i.e. 
its Jacobian is equal to one, %and its the induced kernel being 
invariant to the number of non-zero elements. 
%\jack{I moved this below, need to make sure it is clear here though that we use l1 becuase it is convenient}
%The induced distribution on each $\bm{\Theta}^{\,j}$ is a mixture of a discrete masses at zero and continuous distributions on $\mathbb{R}^{D \times D \times P}$, and as a result, posterior MCMC samples will contain both exact zeros and non-zeros, facilitating the straightforward calculation of posterior inclusion probabilities. {\color{red}(May be in Posterior Sampling?)} \jack{Move below?}

%\jack{Notation reminder: matrix bold upper, element not bold lower, upper case $\beta$ is B - check this}

\subsection{Switching Model with Arbitrary Dwell Distribution}
\label{seq:switching_model}

\textcolor{black}{Popular hidden state models assume that the latent state sequence $\bm{z}$ evolves as a Markov process and corresponds to assuming the dwell time in each state is geometrically distributed. However, there is no reason beyond computational convenience to believe that dwell times in active and resting states are geometric.}
%We further model the hidden states $\bm{z}$ as evolving according to a hidden semi-Markov model (\HSMM) with dwell durations $\{d_j  \}_{j=1}^{K}$. 
\textcolor{black}{We instead model the hidden states $\bm{z}$ as evolving according to a hidden semi-Markov model (\HSMM) with general dwell duration distributions $\{p \, (\, d_j = s \, | \,  \bm{\lambda}^{\,j})\}_{j=1}^{K}$ parametrised by $\{\bm{\lambda}^{\,j}\}_{j=1}^K$ (e.g. the rate and dispersion of a negative binomial distribution).} 
\textcolor{black}{This generalised framework allows us to consider several possible dwell distributions (including the \HMM) and see which is best supported by the data.} 
Following \citet{langrock2011hidden} and \citet{hadj2022bayesian}, 
%a $K$-state \HSMM with dwell durations $\{d_j (r) \}_{j=1}^{K}$ and hidden state sequence $\bm{z} = (z_1, \dots, z_T)$ $\in \{1, \dots, K\}^D$, 
such a \HSMM 
can be approximated by an extended state-space \HMM with latent states $\bm{z}^{\star} = (z^{\star}_1 , \ldots , z^{\star}_T ) \in \{1, 2, \dots, \bar{B} \}$, where $\bar{B} = \sum_{i = 1}^{K} b_i$ and $\bm{b} = (b_1, \dots, b_K)$ are fixed positive integers (\textit{thresholds}). 
Each \HSMM state is mapped to a group of $b_j$ \HMM states that are associated with the same \VAR emission distribution (Eq. \ref{Equ:VAR_emission}), and these group of states is called a \textit{state aggregate} $B_j = \big\{ \,  \bar{b}: \sum_{i=0}^{j-1} b_i < \bar{b} \leq \sum_{i=0}^{j} b_i \,   , b_0 = 0\big\}$.
\textcolor{black}{The \HSMM dwell time $d_j$ then correspond to the number of consecutive time points that the latent state is a member of the state aggregate $B_j$.}
%Each \HSMM state $j$ corresponds to a \textit{state aggregate} $B_j = \big\{ \,  b: \sum_{i=0}^{j-1} b_i < b \leq \sum_{i=0}^{j} b_i \,   , b_0 = 0\big\}$, {\color{red}a group of $b_j$ \HMM states that are associated with the same \VAR emission distribution (Eq. \ref{Equ:VAR_emission}).}  
 The state aggregates allow the \HMM to capture the more complex \HSMM dwell distributions through the special structure of the Markov state transition matrix $\bm{A} = \big\{  A_{il} \big\}_{i,l=1}^{\bar{B}}$, where $A_{il} = p \, (\, z^\star_{\, t} = l \, | \, z^\star_{\, t-1}  = i \, )$ are completely determined by the \HSMM dwell distribution $p \, (\, d_j = s \, | \,  \bm{\lambda}^{\,j})$ (within state aggregate moves) and the between \textcolor{black}{state aggregate transition probabilities $\bm{\pi}^{\,j} = (\pi_{j1}, \dots, \pi_{jK})$, in which  $\pi_{jk} = p \, (\, z_t = k \, | \, z_{t-1}  = j, \, z_t \neq j)$ for $j, k = 1, \dots, K$.} Note that $\pi_{jj} = 0$, since self transitions are prohibited. 
 %\textcolor{red}{Here $\bm{\lambda}^{\,j}$ denotes the parameters modelling the state-specific dwell durations (e.g. the rate and dispersion of a negative binomial distribution)}. 
 See Supplementary Material for the exact form of the matrix $\bm{A}$.

The generative framework modelling the evolution dynamics of the hidden states can then be summarized as
     \begin{equation}
\bm{\pi}^{\,j} \sim \text{Dirichlet}_{K-1}\,  (\bm{\alpha}_0), \quad d_j  \sim g(\bm{\lambda^{\,j}}), \quad \bm{\lambda}^{\,j} \sim G, \qquad j = 1, \dots, K\label{Equ:Dwell_Dist}
    \end{equation}
\begin{equation*}
 z^\star_{\,t} \, | \, z^\star_{\,t-1} \sim \bm{A}_{\,z^\star_{\, t-1}} \quad \, \, \,   t = 1, \dots, T \qquad \bm{A}_i := \psi(\bm{\pi}, \bm{\lambda}), \quad i = 1, \dots, \bar{B},
    \end{equation*}
 %\begin{equation*}
 %z^\star_{\,t} \, | \, z^\star_{\,t-1} \sim \bm{\phi}_{\,z^\star_{\, t-1}} \quad \, \, \, \, \,  t = 1, \dots, T,
%    \end{equation*}
where, we assume that the initial state has distribution $\bm{A}_0 = (A_{01}, \dots, A_{0\bar{B}})$, e.g. $z_0^{\star} 
 \sim \bm{A}_0$, where we follow a common practice and assume a uniform distribution for the initial state \citep{zucchini2017hidden}. Here $\psi(\cdot)$ indicate the mapping from $\bm{\pi}  = \{ \bm{\pi}^{\,j}\}_{j=1}^{K}$ and $\bm{\lambda} = \{ \bm{\lambda}^{\,j} \}_{j=1}^{K} $ to the matrix $\bm{A}$, and  $\bm{A}_i$ denotes the $i^{th}$ row of $\bm{A}$. \citet{langrock2011hidden} showed that this formulation %choice of $\bm{A}$ 
allows for the representation of any duration distribution, and yields an \HMM that is, at least approximately, a reformulation of the underlying \HSMM. 
\cite{hadj2022bayesian} deployed a likelihood approximation method and showed that such a formulation can speed \HSMM inference up considerably while incurring minimal statistical error. 

{\color{black} To provide intuition about the extended state-space \HMM,  Figure \ref{fig:HSMMapprox_vs_HMM_example} illustrates an example of formulating an \HSMM through an \HMM, where for the sake of the illustration, this representation is made exact, namely there is no approximation. A two-state \HSMM characterized by emission distributions $f(\bm{\theta}_j)$, and dwell-duration distributions $d_j(r)$, for  $j=1, 2$, is represented as an \HMM with extended state space $\bar{B} = 7$ with dwell-threshold $\bm{b} = (4, 3)$ for state aggregates $B_1$ and $B_2$, respectively. Note that the dwell-thresholds are selected to cover the entire support of the duration distributions, so that this \HMM formulation corresponds to its exact \HSMM representation. The transition matrix $\bm{A}$ is  completely determined by the \HSMM dwell distribution $p \, (\, d_j = r )$ (within state aggregate moves) and the between state aggregate transition probabilities $\pi_{12}$ and $\pi_{21}$, which in this case are equal to one, since we consider a two-state \HSMM.

 \begin{figure}[htbp]
 \centering
\includegraphics[width=0.75\linewidth]{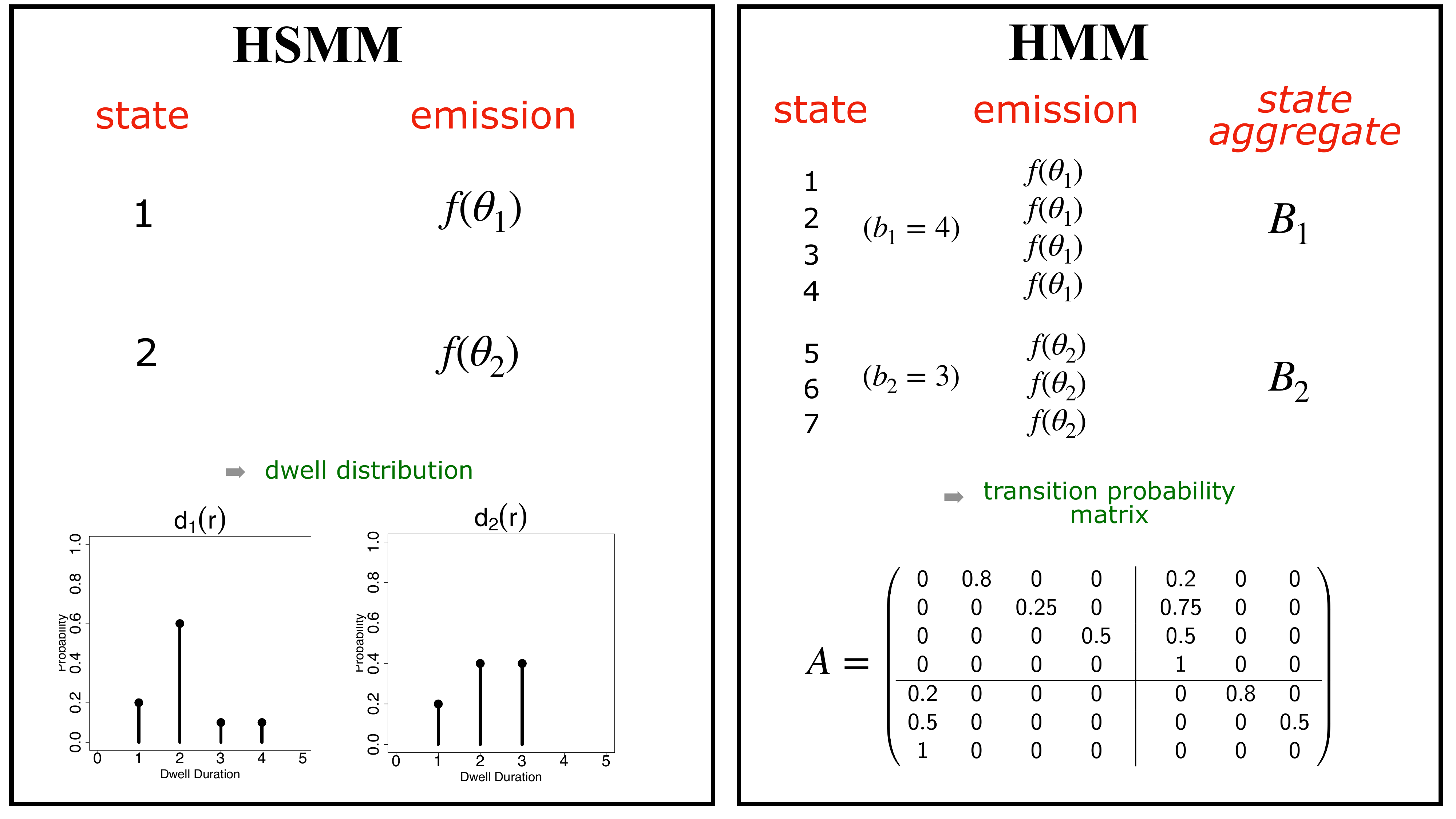}
\caption{Example of formulating an \HSMM through an \HMM.}
\label{fig:HSMMapprox_vs_HMM_example}
\end{figure}
}

\subsubsection{State-dwell parameters and non-local priors}

\begin{figure}[htbp]
\begin{center}
\includegraphics[trim= {0.0cm 0.0cm 0.0cm 0.0cm}, clip,  
width=0.49\columnwidth]{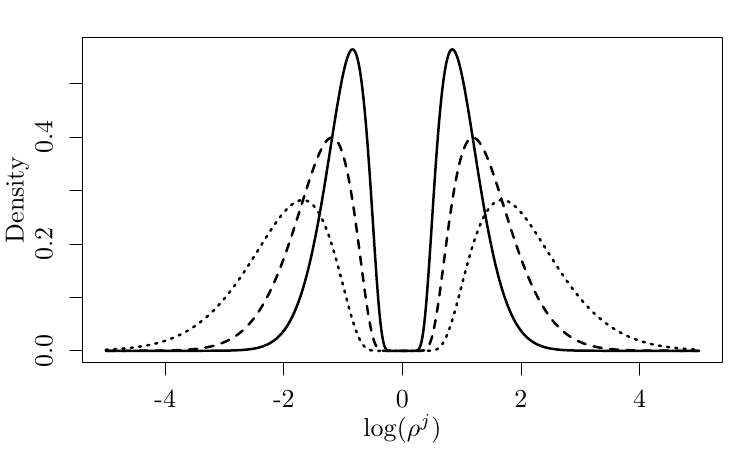}
\includegraphics[trim= {0.0cm 0.0cm 0.0cm 0.0cm}, clip,  
width=0.49\columnwidth]{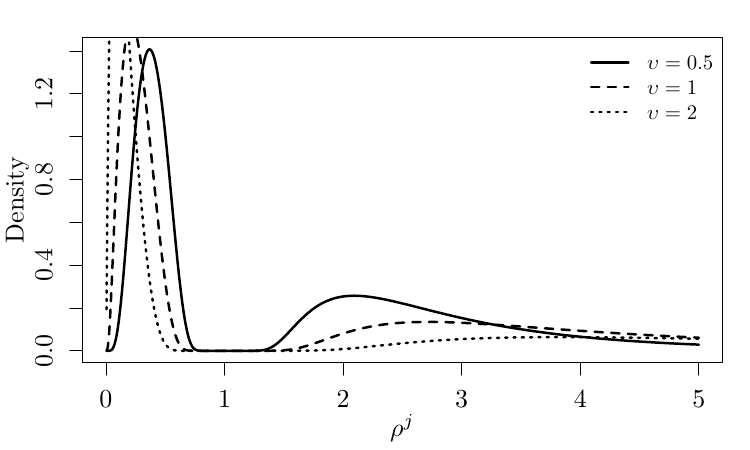}
%trim={<left> <lower> <right> <upper>}
\caption{\textit{The eMOM} non-local prior for $\log \rho^{\, j}$ and the implied prior on $\rho^{\, j}$ for several different vales of the hyperparameter $\upsilon^{\, j}$.}
\label{Fig:MOM_prior}
\end{center}
\end{figure}

Next, we specify the priors for the state dwell parameters $\bm{\lambda}^{\,j}$. We consider two choices: the negative binomial distribution, and the geometric distribution (corresponding to the \HMM). %\jack{and chose in a data driven manner which is better supported}. 
\textcolor{black}{The negative binomial dwell distribution is parameterized by its mean and overdispersion parameter as $\bm{\lambda}^{\,j} = \{m^{\, j}, \rho^{\,j}\}$ and constitutes a parsimonious extension of the geometric distribution, with the ability to capture over and under dispersion with only one extra parameter. }
The probability mass function of a negative-binomial, shifted to have strictly positive support, is given by
\begin{align}
    g^{NB}(d_j; \bm{\lambda}^{\, j}) = \frac{\Gamma(d_j - 1 + \rho^{\, j})}{\Gamma(\rho^{\, j})(d_j - 1)\!}\left(\frac{\rho^{\, j}}{\rho^{\, j} + m^{\, j}}\right)\left(\frac{m^{\, j}}{\rho^{\, j} + \lambda^{\, j}}\right)^{d_j - 1}, \quad d_j = 1, 2, \ldots.
\end{align}
The mean dwell time under negative binomial dwell is $m^{\, j} + 1$ and its variance is $m^{\, j} + \frac{{m^{\, j}}^2}{\rho^{\, j}}$.
 Note that the geometric distribution can be obtained from the negative binomial distribution for $\rho^{\, j} = 1$ and $m^{\, j} = \frac{\tilde{\pi}_{jj}}{1 - \tilde{\pi}_{jj}}$, where $\tilde{\pi}_{jj} := p(z_t = j | z_{t-1} = j)$ denotes the (Markovian) probability of self transitions in the \HMM. Thus, for the negative binomial overdispersion parameters $\rho^{\, j}$, we consider two candidate priors:  (i) \textcolor{black}{a so called `local'-prior (\LP), i.e. $ p^{\,LP}\left(\rho^{\, j}\right) = \mathcal{IG}\left(\rho^{\, j}; c_0^{NB}, c_0^{NB} + 1\right)$ such that the prior mode at $\rho^{\, j} = 1$}, %and assigns strictly non-zero density in the region 
namely the value of $\rho^{\, j}$ that recovers the simpler \HMM; (ii) a `non-local' prior (\NLP) \citep{johnson2012bayesian} that assigns 0 density in the region around $\rho^{\, j} = 1$. Specifically, we place an exponential moment prior (\textit{eMOM}) \citep{rossell2017nonlocal} prior on $\log \rho^{\,j}$,
\begin{align}
    %p^{\,NLP}( \log \rho^{\,j} | \upsilon^{\, j}) = \frac{\left(\log \rho^{\,j}\right)^2}{\tau^{\,j}} \times \mathcal{N}\left(\log \rho^{\,j}; 0, \sqrt{\upsilon^{\, j}}\right),\label{Equ:MOM}
    p^{\,NLP}( \log \rho^{\,j} | \tau^{\, j}) = \exp\left\{\sqrt{2} - \frac{\upsilon^{\,j}}{\left(\log \rho^{\,j}\right)^2}\right\} \times \mathcal{N}\left(\log \rho^{\,j}; 0, \sqrt{\upsilon^{\, j}}\right),\label{Equ:MOM}
\end{align}
%same as putting the prior on log rho^{-1} = -log rho  as its symmertic!
that has 0 density at $\log \rho^{\,j} =  0$. \textcolor{black}{Compared with other \NLP's, the \textit{eMOM} prior has lighter tails allowing for a prior that is non-local but still places high density on reasonable values.} Figure \ref{Fig:MOM_prior} plots the \textit{eMOM} prior for $\log \rho^{\,j}$ for several values of the hyperparameter $\upsilon^{\, j}$ alongside the implied prior for $\rho^{\,j}$. 
%Section \ref{Sec:HMM_sims} illustrates how such a prior can improve Bayesian model selection when the data was generated by the simpler \HMM model. 
For the mean parameter $m^{\, j}$, we specify a Gamma prior, namely  $m^{\, j} \sim \mathcal{G}(a_0^{NB}, b_0^{NB})$.

% Full specifications of their parameterizations are given in Section \ref{Sec:dwell_distributions_app}. {\color{red}(ADD HERE, IN THE PAPER)}\jack{Add forms of dwell in here}

The geometric dwell distribution is parameterized by the \HMM state's probabilities of self transition $\bm{\lambda}^{\,j} = \{\tilde{\pi}_{jj}\}_{j=1}^{K}$, with probability mass function, shifted to have strictly positive support, given by 
\begin{align}
    g^{geom}(d_j; \bm{\lambda}^{\, j}) = \left(\tilde{\pi}_{jj}\right)^{d_j - 1}\left(1 - \tilde{\pi}_{jj}\right), \quad d_j = 1, 2, \ldots,
\end{align} 
with mean dwell time specified as $\frac{1}{1 - \tilde{\pi}_{jj}}$. In this case, the \HMM transition probabilities $\tilde{\bm{\pi}}^{j} = (\tilde{\pi}_{j1}, \dots, \tilde{\pi}_{jK})$ are assigned a Dirichlet prior on the ($K-1$) dimensional simplex, i.e. $\bm{\tilde{\pi}}^{j} \sim \text{Dirichlet}_K(\bm{\delta}^{Geom}_0)$, for $j=1, \dots, K$.

\begin{figure}[htbp]
     \centering
     \includegraphics[width=0.75\linewidth]{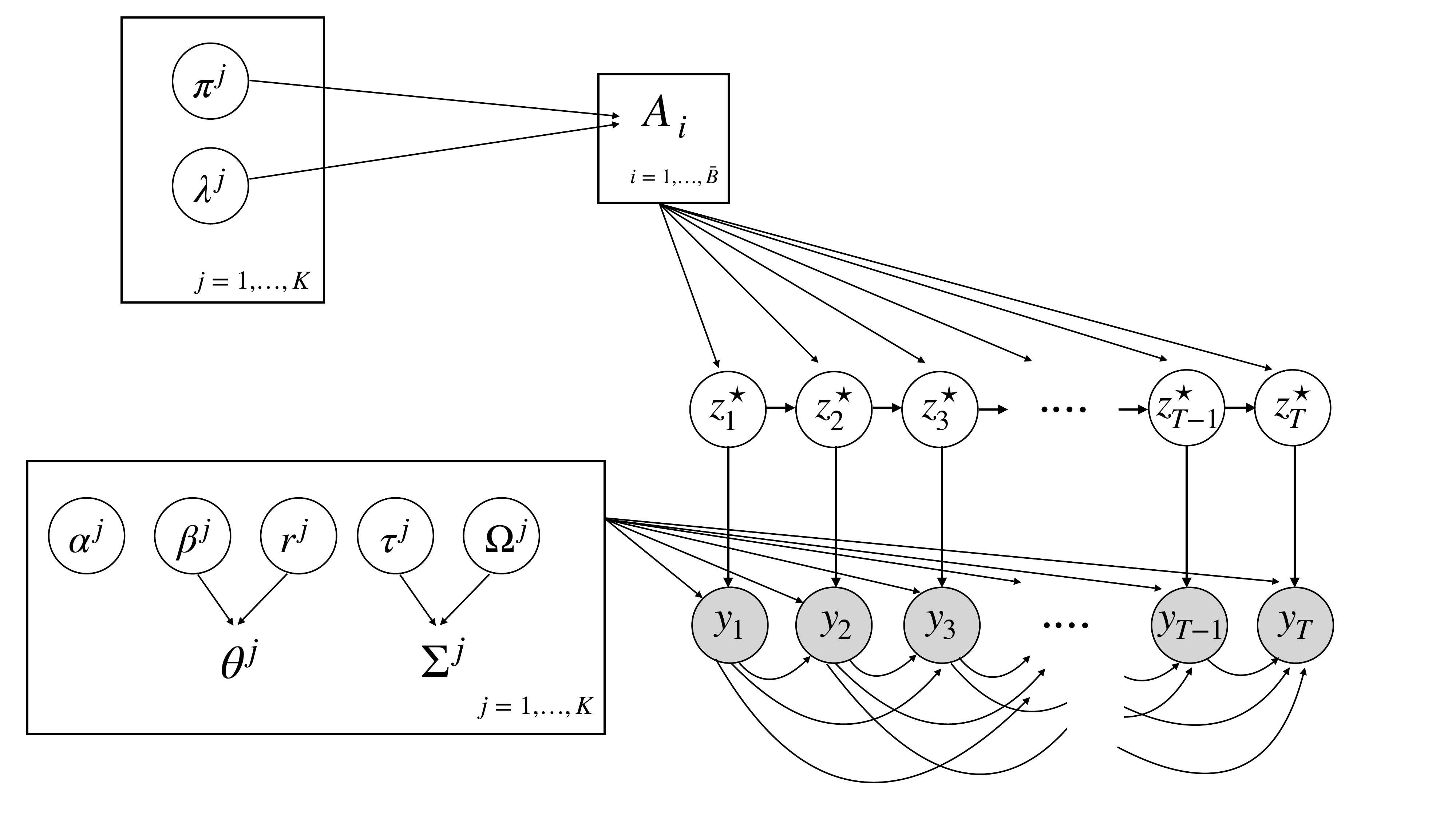}
     \caption{Graphical model representation of the proposed sparse \VAR approximate \HSMMs. \textcolor{black}{Transition probabilities $\bm{A}_i$  are solely determined by $\bm{\pi}_j$ and $ g( \bm{\lambda}_j)$, and thus the probabilities $\bm{A}_i$ are not considered as random variables themselves.} Similarly, while the likelihood depends on $\bm{\Theta}^{\,j}$ and $\bm{\Sigma}^{\,j}$, it is convenient to specify priors for latent parameters $\bm{\beta}^{\, j}$, $r^{\, j}$, $\bm{\tau}^{\, j}$ and $\bm{\Omega}^{\, j}$ for reasons outlined in Sections \ref{seq:VAR_emisison} and \ref{Sec:l1_ball}.}
     \label{Fig:GraphicalModel}
 \end{figure}

Figure \ref{Fig:GraphicalModel} provides a graphical model depicting the structure of the proposed sparse \VAR HSMM. 

\subsection{Posterior Sampling}{\label{Sec:PosteriorSampling}}

\textcolor{black}{Let $\bm{\eta} = \big\{ \, (  \bm{\pi}^{\,j}, \, \bm{\lambda}^{\,j}, \, \bm{\alpha}^{\,j}, \, \bm{\beta}^{\,j}, \, r^{\,j}, \, \bm{\Omega}^{\,j}, \, \bm{\tau}^{\,j}) \, \big\}_{j=1}^{\,K}$ be the parameters of interest, where we recall that the autoregressive matrices $\bm{\Theta}^{\,j}$ are a deterministic transformation of parameters $\bm{\beta}^{\,j}$ and $r^{\,j}$ through the $l_1$-ball projection prior, and the covariance matrices $\bm{\Sigma}^{\,j}$ are constructed deterministically from $\bm{\Omega}^{\,j}$ and $\bm{\tau}^{\,j}$.} \textcolor{black}{Throughout, we denote prior distributions by $p(\cdot)$ and differentiate these from their corresponding posterior distributions $p(\cdot \mid y)$ by the conditioning on $y$. We write the posterior distribution as}

{ \color{black}
\begin{equation}
p \, ( \,  \bm{\eta} \, | \, \bm{y} ) \propto \mathscr{L}\,  ( \bm{y} \, | \,  \bm{\eta} ) \, \times \, \bigg[ \, \prod_{j=1}^{K} \,  p \, ( \bm{\pi}^{\,j} ) \, \times \,   p \, ( \bm{\lambda}^{\,j}  ) \, \times \,   p \, ( \bm{\alpha}^{\,j}  ) \, \times \, p \, ( \bm{\beta}^{\,j}) \,  \times  p \, ( r^{\,j}) \, \times \, p \, ( \bm{\Omega}^{\,j}) \, \times \, p \, (\bm{\tau}^{\,j})  \bigg] \, ,
\label{eq:posterior}
\end{equation}
}
where the likelihood $\mathscr{L}\, ( \, \cdot \, )$ is defined as %(see e.g. \citealt{zucchini2017hidden})
\begin{equation}
\mathscr{L}\,  ( \bm{y} \, | \bm{\eta} ) =  \bm{\pi}_0^{\,\star\,'} \, \bm{P}\,(\bm{y}_1) \,  \bm{A} \,  \bm{P}\,(\bm{y}_2) \, \bm{A} \, \cdots \, \bm{A} \, \bm{P}\,(\bm{y}_{T-1}) \,  \bm{A} \, \bm{P}\,(\bm{y}_{T}) \, \mathbf{1},
\label{eq:likelik}
\end{equation} with the diagonal matrix $\bm{P}\,(\,y\,)$ of dimension $\bar{B} \times \bar{B}$ defined as 
\begin{equation}
\bm{P}\,(\,\bm{y}\,) =  \text{diag} \, \big\{ \, \underbrace{p \,( \bm{y} \, | \, \bm{\alpha}^{\,1}, \, \bm{\Theta}^{\,1}, \bm{\Sigma}^{\,1}), \,  \dots,\, p  \,( \bm{y} \, | \, \bm{\alpha}^{\,1}, \, \bm{\Theta}^{\,1}, \bm{\Sigma}^{\,1})}_{b_1 \, \, \text{times}}, \, \dots, \, \underbrace{p \, ( \bm{y} \, | \, \bm{\alpha}^{\,K}, \, \bm{\Theta}^{\,K}, \bm{\Sigma}^{\,K}) \dots p \, ( \bm{y} \, | \, \bm{\alpha}^{\,K}, \, \bm{\Theta}^{\,K}, \bm{\Sigma}^{\,K})}_{b_K \, \, \text{times}} \big\},
\nonumber%\label{eq:diag_mat}
\end{equation}
and $ p \, ( \bm{y} \, | \, \bm{\alpha}^{\,j}, \, \bm{\Theta}^{\,j}, \bm{\Sigma}^{\,j})$ the probability density of the \VAR emission distributions (Eq. \ref{Equ:VAR_emission}). Here, $\mathbf{1}$ denotes  an $\bar{B}$-dimensional column vector with all entries equal to one  and $\bm{\pi}_0^{\,\star}$ represents the initial distribution for the state aggregates.  
Note that if we assume that the underlying Markov chain is stationary, $\bm{\pi}_0^{\,\star}$ is solely determined by the transition probabilities $\bm{\Phi}$,  i.e. $\bm{\pi}_0^{\,\star} \, =  ( \bm{I} - \bm{\Phi} + \bm{U})^{\,-1} \, \mathbf{1}$, where $\bm{I}$ is the identity matrix and $\bm{U}$ is a square matrix of ones.  Alternatively, it is possible to start from a specified state, namely assuming that $\bm{\pi}_0^{\,\star}$ is an appropriate unit vector, e.g. $(1, 0,  \dots, 0)$, as suggested by \citet{leroux1992maximum}. 
The computation of the likelihood in Eq. \eqref{eq:likelik} is often subject to numerical underflow and hence its practical implementation usually requires appropriate scaling \citep{zucchini2017hidden}. Furthermore, for multivariate likelihoods the matrix multiplications are implemented on the log-scale.

The posterior distribution \eqref{eq:posterior} is not available in closed form, therefore we conduct inference using \MCMC methods. In particular, we use Hamiltonian Monte Carlo (\HMC, \citealt{duane1987hybrid}), which uses a discretisation of the Hamiltonian dynamics to propose a joint parameter update and accepts or rejects this with the usual Metropolis Hasting probability, and the No-U-Turn Sampler (\NUTS) \citep{hoffman2014no}, which automatically tunes the Hamiltonian step size to the geometry of the targeted posterior. The combination of the two allows for efficient exploration of the multivariate parameter space. 
We take advantage of \HMC and \NUTS implementations  available in the probabilistic programming language \stan{} \citep{carpenter2016stan}. All \stan{} requires is that the user write out their model and the computations required to evaluate the likelihood and \stan{} compiles the models and optimises sampling to its geometry. %{\color{red}One sentence here explaining that HMC is an approximation method.} {\color{red}Here need to describe the output from the HMC and how to use it to perform inference, maybe also estimate of the marginal likelihood. }

The \MCMC implementation outputs $N$ samples $\left\{\bm{\eta}^{(i)}\right\}$, for $i=1,\ldots,N$, approximating the posterior distribution. These samples can be used to estimate posterior summaries such as means and variances for different parameters. 
%% OLD %%
\textcolor{black}{A particular feature of the $l_1$-ball prior is that posterior samples for any of the $\theta_{pil}^{\,j}$, transformed from the corresponding samples of $\beta_{pil}^{\,j}$, may contain exact zeros. Therefore, the posterior probability for the presence of a temporal \VAR connection between $i$ and $l$ in lag $p$ and regime $j$ is straightforward to estimate as the frequency of these samples that are non-zero.   We further remark that 
the parameters $\theta_{pil}^{\,j}$ enter into the likelihood as a deterministic transformation of the parameters $\beta_{pil}^{\,j}$
and $r^{\,j}$, which are easy to sample via HMC as they are strictly continuous.} 
%the induced posterior distribution on each $\bm{\Theta}^{\,j}$ is a mixture of a discrete masses at zero and continuous distributions on $\mathbb{R}^{D \times D \times P}$, and as a result, posterior MCMC samples will contain both exact zeros and non-zeros, facilitating the straightforward calculation of posterior inclusion probabilities.
%Further, the posterior samples for any $\theta_{pil}^{\,j}$ may contain exact zeros allowing the posterior probability for the presence of a temporal \VAR connection between $i$ and $l$ in lag $p$ and regime $j$ to be estimated as the frequency of these samples that are non-zero. \jack{does bit about discrete and continuous go here} 
Further, \stan's compatibility with bridge sampling \citep{meng1996simulating,meng2002warp,gronau2017bridgesampling} allows for the use of posterior samples to estimate the marginal likelihood $p \, ( \, \bm{y} \,) := \int \mathscr{L}\,  ( \bm{y} \, | \,  \bm{\eta} ) \, p \, ( \bm{\eta} ) \, d\bm{\eta}$ and conduct Bayesian model selection for the dwell distribution as in \cite{hadj2022bayesian}. Similarly to \cite{hadj2022bayesian}, our implementation is able to take advantage of the sparse nature of the state transition matrix $\bm{A}$ to provide considerable speed up. Section \ref{Sec:l1ball_projection_alg} details some reparametrizations that were used to improve mixing.

%\jack{Add something about the reparametrisations we did to the parameter space to improve sampling, make it look like we didn't just plug into stan }

%\jack{Something about the sparse matrix multiplication and log-Sum-Exp}

\section{Simulation Studies}{\label{Sec:Simulations}}
\subsection{Data generation}
\begin{figure}[ht!]
\centering
\includegraphics[trim= {0.0cm 1.25cm 0.0cm 0.0cm}, clip,width=0.95\linewidth]{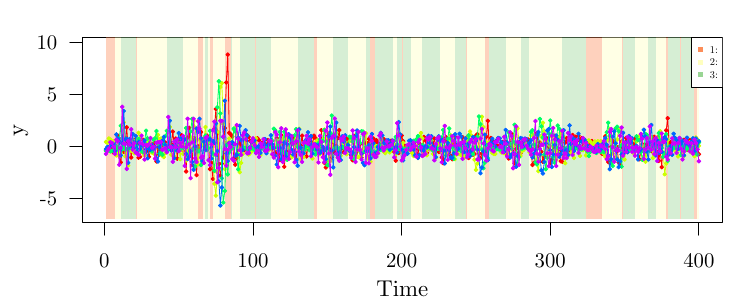}
\includegraphics[trim= {0.0cm 0.0cm 0.0cm 0.5cm}, clip,width=0.95\linewidth]{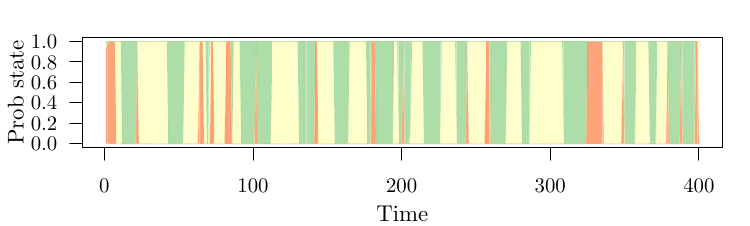}
\caption{Simulation Study. Simulated $D=5$ dimensional time series, plotted on top of one-another, along with the hidden states (top) and state probabilities estimated by the  \HSMM-$l_1$-\NLP (bottom).}
%trim={<left> <lower> <right> <upper>}
\label{fig:Simulation_data}
\end{figure}

To investigate the performance of our model we simulated a $D = 5$-dimensional time series consisting of $T = 400$ observations from a $K=3$ state \HSMM with \VAR emission distributions of order $P=2$ in each regime. The dwell times in each state $j$ were generated from negative binomial distributions $NB(\lambda_j, \rho_j)$ with mean parameters $\lambda_1 = 2$, $\lambda_2 = 10$ and $\lambda_3 = 7$ and overdispersion parameters $\rho_1 = 0.25$, $\rho_2 = 3$ and $\rho_3 = 5$. The between state transition probabilities $\pi_{jk}$ were set to $0.5$, for $j \neq k = 1, 2, 3$. The \VAR parameters were generated such that state 1 was 70\% sparse, state 2 was 30\% sparse and state 3 was 90\% sparse. Non-zero elements of each $\bm{\Theta}^{\, j}_p$ were generated such that they were positive or negative with equal probability and their absolute value was sampled uniformly between 0.2 and 0.8. Additionally, a rejection step was done to check that the sampled values constituted a stable \VAR process. State specific intercept values $\bm{\alpha}^{\,j}$ were sampled uniformly between -4 and 4. Finally, the covariance matrices $\bm{\Sigma}^{\,j}$ were generated as  $\bm{\Sigma} = \bm{P}{\,'} \, \text{diag}(\sigma_1, \dots, \sigma_D) \, \bm{P}$, with $\bm{P}$ being an orthogonal matrix and $\sigma_i$ assumed positive. The matrix $\bm{P}$ was constructed by orthogonalizing a random matrix whose entries were simulated from a standard Normal, while each $\sigma_i$ was uniformly sampled in the interval $[1, 3]$. The data were  standardised such that the time series in each dimension had mean 0 and variance 1. The top plot of Figure \ref{fig:Simulation_data} shows all dimensions of the time series and the true state sequence.

\vspace{0.3cm}

%simulation_VARHMM_HSMM_L1ball.Rmd

\subsection{Parameter settings}
%\subsection{Prior Specification}{\label{Sec:PriorSpecifications}}
%To facilitate Bayesian model selection between the \HMM and \HSMM, \cite{hadj2022bayesian} proposed setting \HSMM prior hyperparameters $a_0^{NB}$ and $b_0^{NB}$ to match the prior mean and variance of the expected dwell time to that of a \HMM
To facilitate Bayesian model selection between the \HMM and \HSMM, \cite{hadj2022bayesian} proposed setting \HSMM prior hyperparameters $a_0^{NB}$ and $b_0^{NB}$ to match the prior mean and variance of the expected dwell time to that of an \HMM \textcolor{black}{with Dirichlet priors on its transition matrix}. 
\textcolor{black}{We elicited the prior mean expected dwell to be 10 with standard deviation 5, providing an uninformative prior and considerable mass away from very short dwell times.} Parameters $\bm{\delta}^{Geom}_0$ and $\bm{\delta}^{NB}_0$ can be elicited based on the expected likelihood of switching from any state to any other state.

%\subsubsection{Non-local prior elicitation}{\label{Sec:NLP _elicitation}}

%Finally, hyperparameters $\upsilon^{\, j}$, $j = 1,\ldots, K$ are set such that the prior density on values of $\rho^{\, j}$ whose dwell distributions are less than a total variation distance of 0.1 from the dwell estimated under the geometric model (\HMM), when the negative binomial mean parameter is matched with that estimated under the geometric, is less than 0.01. This ensures that under the \HSMM, the estimated dwell distribution is forced a prior to be non-negligibly different to the \HMM with very high probability. For full specification see Section \ref{App:NLP_hyperparameter} {\color{red}(add here, in the paper)}.
As for the elicitation of the \textcolor{black}{local and non-local priors, local prior hyperparameter $c_0^{NB}$ was set such that $\frac{1}{\rho^{\,j}} \in (0.25, 4)$ with probability 0.95, expressing high prior probability that the \HSMM is not over or under dispersed by a factor of more than 4}, and non-local hyperparameters $\upsilon^{\, j}$, $j = 1,\ldots, K$ in \eqref{Equ:MOM} are set such that the non-local prior density for $\rho^{\, j}$ is such that $P\big(\rho^{\, j} \in (L_{\rho^{\, j}}, U_{\rho^{\, j}}) \big) < 0.01$ where
\begin{align}
    L_{\rho^{\, j}} :&= \left\lbrace \rho \in (0, 1) : \TVD(g^{Geom}(\,\cdot\,; \hat{\pi}_{jj}),\, g^{NB}(\,\cdot\,; \frac{\hat{\pi}_{jj}}{1-\hat{\pi}_{jj}}, \rho) = 0.1\right\rbrace\nonumber\\
    U_{\rho^{\, j}} :&= \left\lbrace \rho \in (1, \infty) : \TVD(g^{Geom}(\,\cdot\,; \hat{\pi}_{jj}),\, g^{NB}(\,\cdot\,; \frac{\hat{\pi}_{jj}}{1-\hat{\pi}_{jj}}, \rho) = 0.1\right\rbrace,\nonumber
\end{align}
$\hat{\pi}_{jj}$ is the posterior mean of the probability of self transition for state $j$ estimated under the \HMM model, $m^{\, j} = \frac{\hat{\pi}^{\, j}_{jj}}{1-\hat{\pi}_{jj}}$ is the parameter of the negative binomial distribution that recovers the geometric distribution when $\rho = 1$ and $\TVD(f, g) = \frac{1}{2}\sum_{d=1}^{\infty} |f(d) - g(d)|$ is the total variation distance between discrete distributions $f$ and $g$. For each state $j$, negligible prior density ($< 0.01$), is placed on regions of $\rho^{\, j}$ whose corresponding negative binomial dwell distribution is less than 0.1 in \TVD from the posterior mean estimated geometric dwell distribution. This ensures that under the \HSMM, the estimated dwell distribution is forced a priori to be non-negligibly different to the \HMM with very high probability.

%\subsubsection{$l_1$-ball prior elicitation}{\label{Sec:l1_prior_elicitation}}

Finally, to complete the $l_1$-ball prior specification, we must elicit values for prior hyperparameters $\sigma_\beta^{\,j}$ and $a_r^{\,j}$. 
In principle, the prior on the latent $\beta^{\,j}_{\,pil}$'s need only be continuous. However, the double-exponential prior in Eq. \eqref{eq:l1_ball_1} is particularly convenient as it allows us to separate the prior specification for $\bm{B}^{\,j}_{\,p}$ from the prior specification on $r^{\,j}$. In particular, \cite{xu2020bayesian} show that prior specification of Eq. \eqref{eq:l1_ball_1} and \eqref{eq:l1_ball_2} results in the implied prior on non-zero $\theta^{\,j}_{pil}$'s being
\begin{align}
    \pi(\theta^{\,j}_{pil} | \theta^{\,j}_{pil}\neq 0) = \frac{1}{2\sigma_{\beta}^{\,j}}\exp\left(-\frac{|\theta^{\,j}_{pil}|}{\sigma_{\beta}^{\,j}}\right),\nonumber
\end{align}
which is the same double-exponential prior placed on the $\beta^{\,j}_{\,pil}$'s. % , resembling a `memory-less' type property. 
Such a property allows us to consider drawing directly from the implied prior distribution for $\theta^{\,j}_{pil}$ with desired prior expected sparsity $s^{\,j} \in (0, 1)$ as
\begin{align}
    \theta^{\,j}_{pil} \sim s^{\,j} \times 0 + (1-s^{\,j})\times\text{DExp}(0, \sigma_{\beta}).\label{Equ:Theta_Sparsity_sim}
\end{align}
We further assume \textit{a priori} that the \VAR process of each state $j$ is \textit{stable} \citep{lutkepohl2005new}, requiring that the reverse characteristic polynomial has no roots in the complex unit circle, i.e.  $\det \big( \bm{I}_D - \bm{\Theta}_1^{\,j}z - \dots - \bm{\Theta}_p^{\,j}z^{p} \big) \neq 0$, for $ |z| \leq 1$, with $z \in \mathbb{C}$, where $\bm{I}_D$ denotes  the identity matrix. %{\color{red}Do we need to say this? I do not think our prior ensures this.}}
%Given our assumption that the \VAR processes in each regime are `\textit{stable}' (Section \ref{seq:VAR_emisison}),
As a result, we elicit $\sigma_{\beta}^{\,j}$ to the maximal value, namely the least informative prior, that maintains a 0.95 probability that the corresponding \VAR coefficients $\bm{\Theta}^{\,j}$ generated according to Eq. \eqref{Equ:Theta_Sparsity_sim} are stable. 
Once $\sigma_{\beta}^{\,j}$ has been elicited, we return to the view of $\bm{\Theta}^{\,j}$ as a function of $\bm{B}^{\,j}$ and $r^{\, j}$,
and set $a_{r}^{\,j}$ such that the prior expected sparsity of the $\bm{\Theta}^{\,j}$'s is $s^{\,j}$, namely that 
\begin{align}
   \mathbb{E}_{r \sim \text{Exp}(a_r^{\,j})}\mathbb{E}_{\bm{\beta}^{\,j}_{p}\sim \text{DExp}(0, \sigma_{\beta}^{\,j})}\left[\frac{1}{D^2P}\sum_{p=1}^{P}\sum_{i=1}^{D}\sum_{l=1}^{D} \mathbb{I}\left(\{\theta^{\,j}_{pil} = 0 \right)\right] = s^{\,j} \nonumber
\end{align}
\textcolor{black}{We set $s^{\,j} = 0.75$ for all $j$ enforcing a large degree of sparsity while still allowing for active connections when supported by the data.}
This two step procedure can be operationalised using Monte Carlo to estimate the probability of stability and the expected sparsity, and one-dimensional grid searches for values of $\sigma_{\beta}^{\,j}$ and $a_{r}^{\,j}$ satisfying the above criteria. The intercepts $\bm{\alpha}^{\, j}$ are given the same prior as the elicited one for $\beta^{\,j}_{\,pil}$'s.

%\jack{This could be simplified by defining an l1-bal function}

\subsection{Results}{\label{Sec:HSMM_sims}}
%We consider investigating samples from the posterior of the four following models:
We considered the  following \textcolor{black}{competing} models, \textcolor{black}{including our proposed approach}:
\begin{itemize}[itemindent=0pt,leftmargin=*]
\itemsep-0.3em 
    \item[(i)] The \VAR-\HMM model, assuming geometric dwell distributions, with $l_1$-ball prior for the \VAR coefficients (\HMM-$l_1$).
    \item[(ii)] The \VAR-\HSMM model with negative binomial dwell distribution, local prior for $\rho^{\,j}$ and $l_1$-ball prior for the \VAR coefficients (\HSMM-$l_1$).
    \item[(iii)] The \VAR-\HSMM model with negative binomial dwell distribution, non-local prior for $\rho{\,^j}$ and double exponential shrinkage prior, i.e. \LASSO, \citep{park2008bayesian} for the \VAR coefficients (\HSMM-\LASSO-\NLP).
    \item[(iv)] \textcolor{black}{The proposed appraoch}, namely the \VAR-\HSMM model with negative binomial dwell distribution, non-local prior for $\rho{\,^j}$ and $l_1$-ball prior for the \VAR coefficients  (\HSMM-$l_1$-\NLP).
    \textcolor{black}{\item[(v)]  Following a similar approach to \citet{allen2014tracking}, we utilize a sliding window to estimate time-varying \VAR coefficient matrices via maximum likelihood, as features to be used in a $k$-means clustering algorithm (with $k$ set equal to three and two for the simulation study and application, respectively); in such a way, we obtain an estimate of the hidden state sequence.  Finally, state-specific \VAR matrices were reestimated using the observations assigned to each distinct state. We refer to this frequentist procedure as \texttt{VARslide}.}
\end{itemize}

We assumed the prior specifications detailed as above, where the prior scale of the double exponential (\LASSO) shrinkage prior in \HSMM-\LASSO-\NLP was estimated assuming zero sparsity. The \HSMM models are approximated using $\bm{b} = (15, 15, 15)$ which a posterior diagnostic (Figure A.3, Supplementary Material) indicated was big enough to imply negligible dwell approximation. For each Bayesian model, our procedure was run for 5,000 iterations, 1,000 of which were discarded as burn-in, \textcolor{black}{while for \VARslide, the size of the sliding window was selected in such a way to maximize model
selection performances averaged over the different states}.
%For each proposed model we sampled a 6,000 iteration \MCMC chain, 1,000 of which are discarded as burn-in to approximate their posteriors.

\begin{table}[ht]
\centering
\caption{Simulation Study. Estimated log-marginal likelihood (log-ml), state classification accuracy (acc) of most likely state sequence, Brier Score (BS) for state estimation, mean absolute error (MAE) of the posterior mean \VAR parameter estimation and Brier Score for parameter selection under the five models considered.}
\begin{tabular}{lccccc}
  \hline
  Method & log-ml & acc & state - BS & \VAR - MAE & \VAR - BS \\ 
  \hline

  \HMM-$l_1$ & -113.818 &    1.0 & 3.3058$\times 10^{-4}$ & 1.7334$\times 10^{-2}$ & 8.3493$\times 10^{-2}$ \\ 
 \HSMM-$l_1$ & -119.651 &    1.0 & 1.1448$\times 10^{-4}$ & 1.7369$\times 10^{-2}$ & 8.3745$\times 10^{-2}$ \\ 

 \HSMM-\LASSO-\NLP & -215.695 &    1.0 & 1.6205$\times 10^{-4}$ & 2.7277$\times 10^{-2}$ & 6.2000$\times 10^{-1}$\\
  \HSMM-$l_1$-\NLP & \textbf{-113.136} &    1.0 & \textbf{1.0462$\mathbf{\times 10^{-4}}$} & \textbf{1.7220$\mathbf{\times 10^{-2}}$} & \textbf{8.1874$\mathbf{\times 10^{-2}}$} \\

   \hline
 \end{tabular}
 \label{Tab:SimResults}
\end{table}
%\jack{check table}

Table \ref{Tab:SimResults} presents a summary of the results for the \textcolor{black}{models (i)-(iv)} models estimated to the simulated data, where we report log marginal likelihood, state accuracy (with most likely state sequence estimated through the Viterbi algorithm, \citealt{zucchini2017hidden}), state Brier score \citep{brier1950verification},  mean absolute error (MAE) of the \VAR coefficients (averaged across $P$ and $K$), and Brier score of the \VAR coefficients. \textcolor{black}{For the frequentist competitor model (v), \VARslide, we report only accuracy (0.61) and MAE ($5.81 \times 10^{-1}$), as the other metrics cannot be computed.} The \HSMM-$l_1$-\NLP achieves the highest marginal likelihood. However, somewhat surprisingly the simpler yet mispecified \HMM-$l_1$ model achieves a higher marginal likelihood than the more complex but correctly specified \HSMM-$l_1$. This is likely a result of considering a relatively short time series resulting in few state transitions and the relatively uninformative \LP placed on all $\rho^{\,j}$. In terms of state estimation, the most likely state estimated by the Viterbi algorithm for each model correctly identified the generating state sequence but the estimated state probabilities under the \HSMM models can be seen to capture the behaviour of the hidden states better by achieving a small Brier Score. For the \VAR parameter estimates, we see that conducting variable selection with the $l_1$-ball achieves a smaller posterior MAE than using a shrinkage prior and that the estimated inclusion probabilities were excellent according to their Brier Score. \textcolor{black}{\VARslide appears to be  by far the worst, as illustrated by low accuracy and high MAE}. Overall, the \HSMM-$l_1$-\NLP is the preferred model considering all the different metrics.

\begin{figure}[ht!]
\centering
\includegraphics[width=0.32\linewidth]{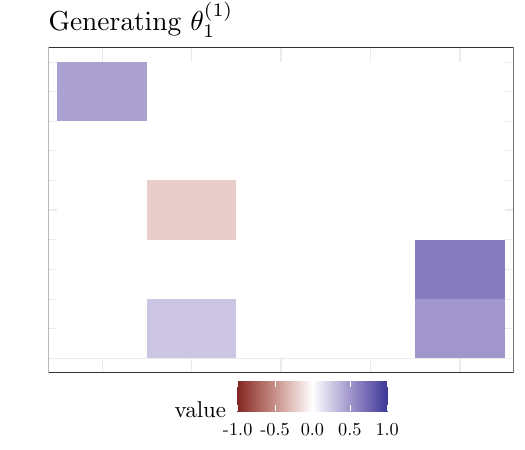}
\includegraphics[width=0.32\linewidth]{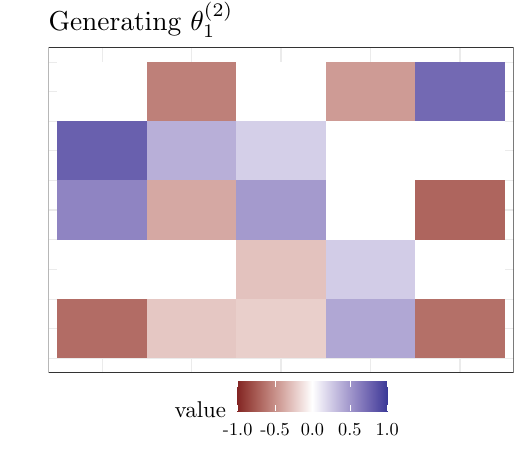}
\includegraphics[width=0.32\linewidth]{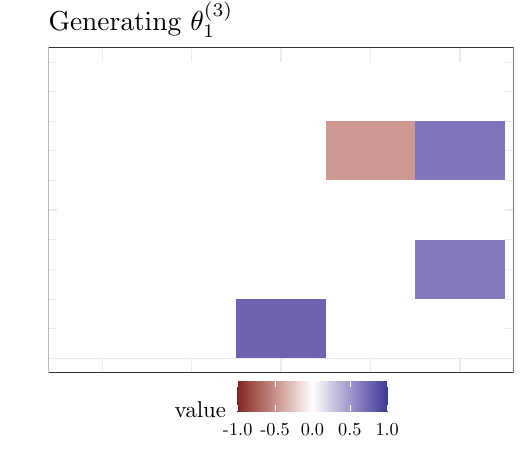}\\
\includegraphics[width=0.32\linewidth]{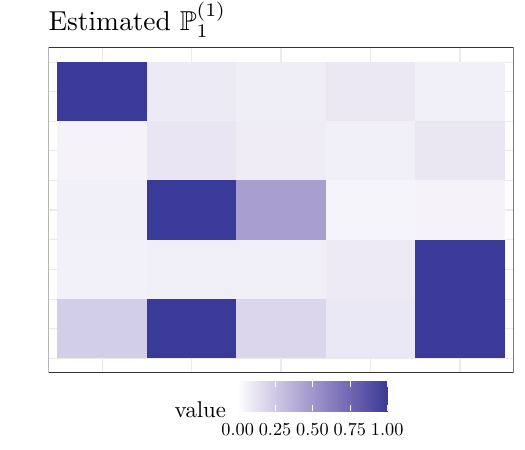}
\includegraphics[width=0.32\linewidth]{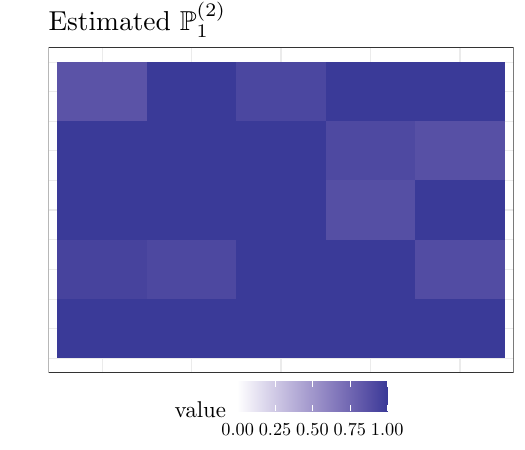}
\includegraphics[width=0.32\linewidth]{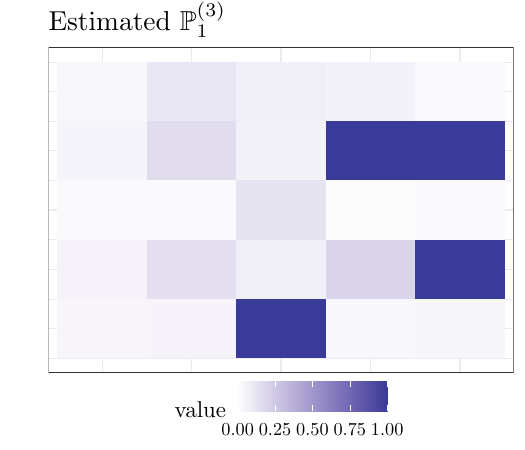}\\
\caption{Simulation Study. Generating \VAR parameters (order 1) and their estimated posterior probability of inclusion.}
\label{fig:Simulation_VAR_inclusion_order1}
\end{figure}

We further investigate the \HSMM-$l_1$-\NLP results. In Figure \ref{fig:Simulation_data} (bottom) we conduct local decoding \citep{zucchini2017hidden} of the hidden state at time $t$, that is we display the estimated time varying probabilities  $p(z_t = j \, | \, \bm{y}, \cdot)$, which show an excellent match with the true generating sequence. Figure \ref{fig:Simulation_VAR_inclusion_order1} compares the estimated inclusion probabilities (for the order 1 \VAR coefficients), in each regime, to their generating values. Firstly, we see that whenever a generating parameter was non-zero, the estimated probability of inclusion for that parameter was 1 or very close to 1. Further, generally when the estimated parameter was zero, the probability of inclusion for that parameter was away from 1 and generally very small. The exception to this is State 2, where the inclusion probabilities of truly 0 parameter are estimated to be close to 1. Interestingly, this state was not particularly sparse, having only 8 in 25 zero parameters. This suggests there may be issues with the $l_1$-ball's performance when the underlying object is not truly sparse. That being said, the $l_1$-ball prior still achieves more accurate posterior mean point estimates in this state than when only using a \LASSO shrinkage prior. 

Similar behaviour was demonstrated for the order 2 \VAR coefficients in Figure \ref{fig:Simulation_VAR_inclusion_order2}. Figure \ref{fig:Simulation_Omega} further demonstrates that the posterior mean estimates values of correlation matrices $\bm{\Omega}^{\, j}$ are accurate to the their generating values and Figure \ref{fig:Simulation_Dwell} plots the posterior predictive for the dwell distribution in each state under the \HMM (geometric dwell) and the \HSMM with \NLP. %Figure \ref{fig:Simulation_Dwell} also compares \HSMM approximation to  the exact dwell distribution..\jack{with what?}

%simulation_VARHMM_HSMM_L1ball.Rmd

%\subsection{Nested Dwell Selection}{\label{Sec:HMM_sims}}

Additional simulation results in Section \ref{Sec:Add_Sims} of the Supplementary Material investigate selecting between a \HMM and \HSMM under the local and non-local priors when the data was generated form the simpler \HMM. These results show that while greater evidence is found in favour of the simpler model when using the non-local prior, the complexity penalisation provided by the local prior appears to be sufficient to always select the correct model in this setting.

%\newpage
\section{Gesture Phases Classification Using Sensor Data}{\label{Sec:GesturePhase}}

% \jack{Ideally we should report the ESS's and $R$-hats from stan, and have MCMC diagnostics in the supplementary }

\subsection{Background information and data processing}

We consider multivariate time series data that arise from a study on human gesture phase segmentation based on sensor data. Recent developments in new sensing technologies and modern detecting devices, as well as increased progress in processing capability, have significantly widened the research horizon for automatic gesture analysis and human-computer interaction \citep{mitra2007gesture, parvathy2021development}. Human gesture analysis seeks to automate tasks relevant to discourse analysis.  Phase segmentation, in particular, aims at identifying and characterising the principal units, hidden states of practical relevance, and dynamical patterns involved in the gesture movement associated with these states \citep{moni2009hmm, wagner2014gesture}. 

The dataset investigated in this article is composed by two different videos recorded using an Xbox Microsoft Kinect\texttrademark  $\,$sensor, where the same user was asked to read two distinct comic strips and to tell the stories in front of the sensor \citep{madeo2013gesture}. As a byproduct of the Kinect\texttrademark  $\,$ device, scalar velocity and acceleration recordings were collected from the left hand (LF), right hand (RH), left wrist (LW), and right wrist (RW) at discrete time intervals known as frames. Measurements were obtained after normalizing the positions of the hand and the wrists according to the position of head and spine. %Thus, a video can be represented by a sequence of frames as input to a segmentation procedure, aiming at identifying rest and gesture positions.  
The data were obtained from the UCI public repository
%\href{https://archive.ics.uci.edu/ml/datasets/gesture+phase+segmentation}{https://archive.ics.uci.edu/ml/datasets/gesture+phase+segmentation} 
\href{https://archive.ics.uci.edu/ml}{https://archive.ics.uci.edu/ml} (\textit{Gesture Phase Segmentation DataSet}) where we used the processed version of the file provided. The same dataset was used by \citet{madeo2013gesture} to segment gestures from rest positions
%where they model the problem as a classification task, 
via the use of Support Vector Machines (SVM). 
%and the application of several pre-processing procedures for extract time-domain and frequency-domain features. 
However, their proposed approach aimed only at segmenting the gesture streams of data, without actually characeterizing the temporal dynamics relative to the different measurements and states. While we aim to model this multivariate time series dataset to identify periods of rest and active gesturing,  we also intend to characterize the existence of switching and  temporal dependences among acceleration and velocity measurements, whilst describing the temporal dynamics underlying these processes.   
%Under our proposed model, changes in physical activity are expressed through switching of the states as well as via autoregressive patterns characterizing each different regime. The temporal connection of these complex multivariate data comprises of not only the serial dependence within each univariate series, but also the interdependence across distinct physiological measurements. 
Furthermore, we use our proposed approach to successfully characterise and predict a new video, corresponding to a different, unseen, story.

 \begin{figure}[ht!]%[htbp]
     \centering
     \includegraphics[width=1.0\linewidth]{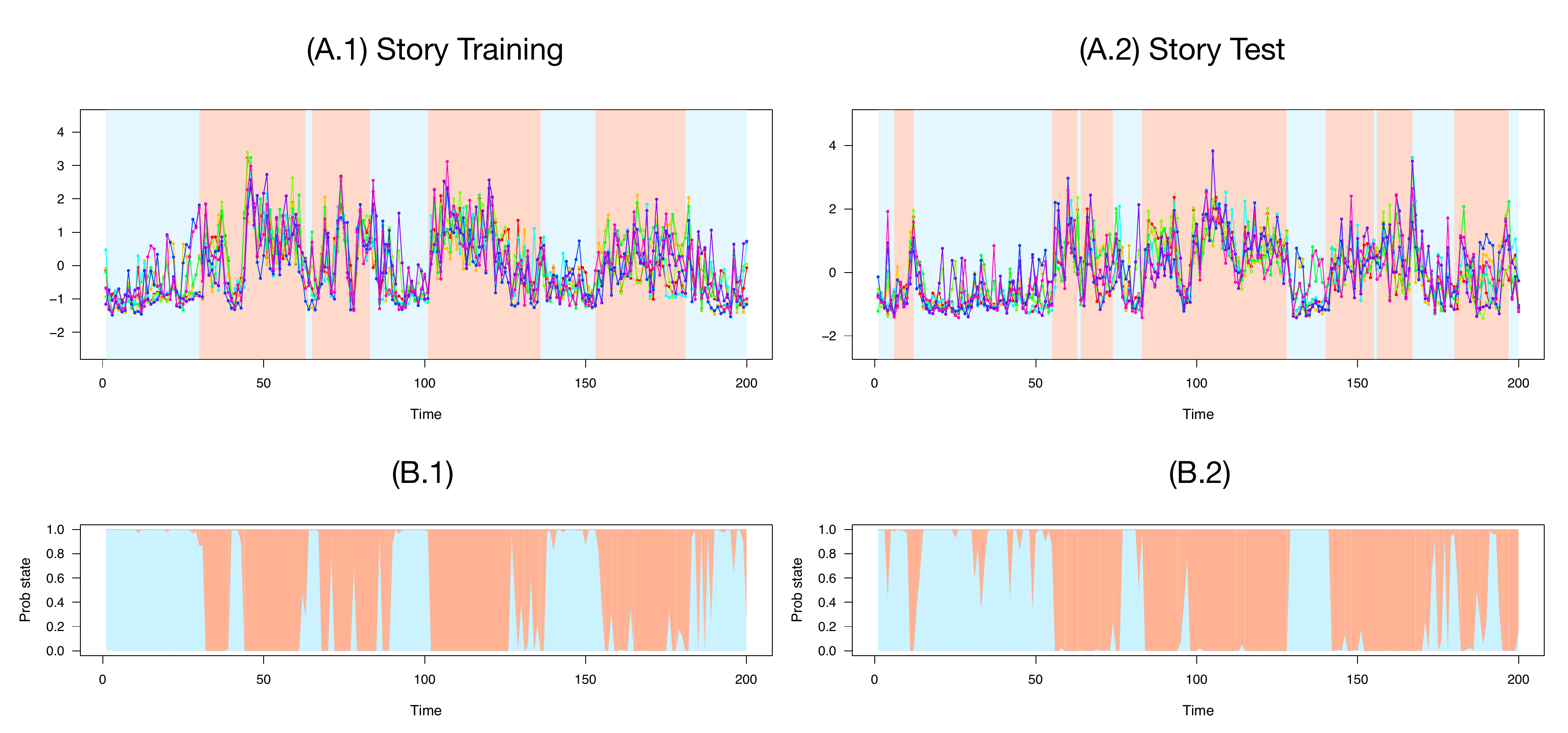}
     \caption{Gesture phase application. Training (A.1) and test (A.2) datasets. Blue vertical bands denote the true resting states, whereas red bands indicate true active states. Time varying state probabilities plots (B.1 and B.2) estimated by our proposed model, \VAR-\HSMM (including $l_1$-ball and \NLP), are also reported for both training and test stories. Our unsupervised model appears to successfully identify regime shifts between the
rest position and gesture activities in a probabilistic way.}
     \label{fig:data_training_vs_test}
 \end{figure}

The time series we analyzed as our training dataset is shown in Figure \ref{fig:data_training_vs_test} (A.1), and the story used as the test dataset is displayed in Figure \ref{fig:data_training_vs_test} (A.2), where individual plots of each dimension are displayed in the Supplementary Material.  They both consist of a stream of gestures resulting in $T=200$ data points and $D = 8$ dimensions. Here, we pre-processed the original data employing the following steps: (i) smoothing, by applying a 2-point moving average filter; (ii) downsampling, by uniformly including every 5 points \citep{pfenninger2017dealing, romanuke2021time}, in such a way that we are able to consider longer time lags using fewer parameters, as well as decreasing the computational complexity of our model;  (iii) transforming, through the square root function \citep{huang2018hidden, hadj2022bayesian} to accommodate Gaussian emissions; (iv) standardizing,   where we have
 scaled the time series such that each dimension $d = 1, \ldots, D$ had mean 0 and variance 1. In this application we have considered the first two hundreds data points of the down-sampled dataset, for both training and test set.  Gestures and rest events were manually segmented by a specialist, providing a ground truth for classification. It is clear that these physiological time series measured on different body regions during the execution of a discourse might exhibit abrupt changes in their structure as an individual experiences regime shifts between the rest position and gesture activities. 

\subsection{Parameter settings}
We consider the different models investigated in the simulation study, considering autoregressive order $P=1$.  \textcolor{black}{For the Bayesian models},  we use the log-marginal likelihood (log-ml) to conduct model selection for the dwell distribution of the hidden state models and evaluate the model's predictive accuracy for the unseen hidden states in training and on a unseen test story,for  For each Bayesian model we used \stan{} to sample 6,000 \MCMC samples from their corresponding posterior distribution, discarding 1,000 for burn-in. The reparametrizations outlined in Section \ref{Sec:l1ball_projection_alg} ensured that the average effective sample size of the remaining samples for all models was in excess of 1000.
%and the minimum was \jack{???}. 
 Figure \ref{fig:Simulation_Dwell_Gesture} also provides a similar diagnostic used in the Section \ref{Sec:HSMM_sims} to demonstrate that the dwell approximation threshold $\bm{b}$ was chosen large enough to well approximate the underlying \HSMM.
%\jack{For each proposed model our Bayesian procedure is run for 6,000 iterations, 1,000 of which are discarded as burn-in. Maybe here describe a bit of the hyperparameters for the different model, maybe you can refer to Section \ref{Sec:Simulations}. Maybe you want to mention convergence of the results, e.g. posterior predictive plot in the Supplementary, e.g. supplementary Material }.  

% \jack{Should we say something about selection of $P$} \benni{Good question, we could run P = 2 and show (hope) its marginal likelihood was less? Regarding the hidden states, we fix $K=2$ since we want only to select between rest and active, in this segmentation application.}.

\subsection{Results}
As a measure of state selection performance, we computed accuracy, sensitivity, specificity, $F1$-score, and Matthew correlation coefficient (MCC) (see e.g. \citet{chicco2020advantages} for an excellent review about these classification measures); to evaluate prediction accuracy for the test set we report the predictive log-likelihood. 
\textcolor{black}{Table \ref{tab:casestudy_classification_states} presents results for our proposed approach and Bayesian competitors, where results for the frequentist approach, (\VARslide) in the training set were accuracy 0.597, sensitivity 0.604, F1 0.645, and MCC 0.191, with a computational time of 8 seconds (we note that we were not able to compute those metrics for the test set since \VARslide does not provide an automatic framework for prediction).} The estimated log-ml demonstrates that there is greater evidence in the training data for the negative-binomial \HSMM compared to the \HMM and that similarly to the simulations, adopting a \NLP allows for greater differentiation between the \HSMM and \HMM. The log-ml further supports inducing sparsity into the \VAR coefficients using the $l_1$-ball over shrinking the coefficients using a Bayesian \LASSO type prior. With regards to the classification of the unseen `true' states, the \HSMM models achieve better state classification than the \HMM and the \NLP further improves the performance of the \HSMM. 
\textcolor{black}{While the \HSMM-$l_1$-\NLP outperforms the \HMM-$l_1$ in terms of state classification on the training data, their performance on the testing data is more similar. We do not, however, believe this is a result of our model over-fitting to the training set, as the model is unsupervised and never sees the class labels. The improved test set log-likelihood is a further indication that the \HSMM better fits the data even though its classification point estimates are similar.}
\textcolor{black}{While \VARslide is significantly faster than the other Bayesian approaches, classification performances for this data appear inferior to our proposed method and the other competitors. Section \ref{Sec:AdditionalGesturePhase} contains plots demonstrating that the \HSMM-\NLP model improves classification accuracy by fitting a heavier-tailed dwell distribution allowing it to estimate longer dwell times.} 
Interestingly, we see that imposing sparsity further helps the hidden state models differentiate between states when compared with the \LASSO shrinkage prior. \textcolor{black}{The improved inference of the \HSMM and incorporating variable selection with the $l_1$-ball does come at a computational cost. Sampling from the posterior of the \HSMM models is about 20 times slower than the \HMM model and incorporating the $l_1$-ball prior is two times slower than the shrinkage prior alternative. However we argue that the computational time is justified by the improved inference and is considerably faster than full \HSMM inference \citep{hadj2022bayesian}}.
%\textcolor{red}{While the results of \cite{hadj2022bayesian} suggest that the computational time required for the \HSMM models here is much lower than their the exact counterparts, sampling from the posterior of the \HSMM models still take around 20 times as long as the \HMM model and incorporating the $l_1$-ball prior takes twice as long as the shrinkage prior alternative.}
%\textcolor{red}{Further, the \NLP improves the reate of selection of the computationally more convenient \HMM would be selected if it was sufficiently supported by the data.}
%Further,  the \HMM model is much faster. The \NLP prior stops the \HSMM from looking like an \HMM and therefore ensures the the extra computational cost at least results in a substantially different model and that the computationally convenient \HMM would be selected if it was sufficiently supported by the data.}

\begin{table}[ht!]
%\small
\centering
%\caption{Gesture phase application. Accuracy, sensitivity, specificity, F1 score, Matthew correlation coefficient (MCC), and estimated log-marginal likelihood (log-ml), for both training and test dataset, where we compare four different models: (i) \VAR-\HMM (with $l_1$-ball); (ii) \VAR-\HSMM (with $l_1$-ball); (iii) \VAR-\HSMM (with \NLP); (iv) \VAR-\HSMM (with $l_1$-ball and \NLP). Run times (in seconds) of the \MCMC algorithm for the training story is also reported.}
\caption{Accuracy, sensitivity, specificity, F1 score, Matthew correlation coefficient (MCC), and estimated log-marginal likelihood (log-ml), for both training and test dataset, for models (i)-(iv). Testing set log-ml is a one step ahead log predictive density. Run times (in seconds) of the \MCMC algorithm for the training story is also reported.}
\begin{tabular}{clccccccc}
  %\hspace{-0.2cm}\\   
  \hline
                                & \multicolumn{1}{c}{} & acc & sens & spec & F1 & MCC & log-ml & time \\ \cmidrule{2-9}
\multirow{4}{*}{Training} 
& \HMM-$l_1$           & 0.835  & 0.860   & 0.802   & 0.856 & 0.662  & -1274.55     & 1557 \\
                                & \HSMM-$l_1$       & 0.850  & 0.877   & 0.814  & 0.870 & 0.693  & -1272.68     & 34498   \\
                                & \HSMM-\LASSO-\NLP       & 0.760  & 0.903   & 0.570  & 0.811 & 0.511  & -1280.48     & 14946   \\
                                & \HSMM-$l_1$-\NLP  & 0.890  & 0.903  & 0.872  & 0.903 & 0.776  & \textbf{-1266.81}     &   24564  \\ \cmidrule{2-9}
\multirow{4}{*}{Test}     
& \HMM-$l_1$         & 0.860  & 0.920   & 0.784   & 0.880 & 0.716  & -986.38     & -   \\
                                & \HSMM-$l_1$      & 0.860  & 0.929   & 0.772   & 0.881 & 0.717  & -980.24     & -   \\
                                & \HSMM-\LASSO-\NLP       & 0.725  & 0.902    & 0.500   & 0.786 & 0.447  & -1020.91     & -   \\
                                & \HSMM-$l_1$-\NLP  &  0.860  & 0.910   & 0.796   & 0.879 & 0.715  & \textbf{-945.13}     & -  \\ 
                             \hline 
%\normalsize
\end{tabular}
\label{tab:casestudy_classification_states}
\end{table}

Table \ref{tab:casestudy_classification_states} also presents the posterior predictive performance of the models (i)-(iv) estimated on the training data to an unseen testing story. The \HSMM provides an improved fit to the unseen data, particularly improving the one step ahead log predictive density, and introducing sparsity also improved predictive performance. Figure \ref{fig:data_training_vs_test} \textcolor{black}{(B.1, B.2)} displays time varying state probabilities estimated by the \HSMM-\NLP model for both training and test stories, showing that our unsupervised model appears to successfully identify regime shifts between the rest position and gesture activities in a probabilistic manner. Section \ref{Sec:AdditionalGesturePhase} in Supplementary Material contains 
%some representative trace-plots and 
a graphical posterior predictive check consisting of the observations alongside 100 draws from the estimated posterior predictive \citep{gelman2006data}.%\jack{should we say something about how we estimate log marignal likelihood in this case, it's different than the training isn'it it? It's like predictive log score something like that, as in \citet{hadj2022bayesian}}. 

%\jack{Describe Table \ref{tab:casestudy_classification_states}, once we have the different models and comment something why we choose \VAR-\HSMM (including $l_1$-ball and \NLP). I mean Bayes factor is super good for the final model, for both training and test set. Maybe here is also a good place to talk about state probabilities plot in Figure \ref{fig:data_training_vs_test} (C.1, C.2), e.g. where we .}

% \begin{table}[htbp]
% \centering
% \begin{tabular}{lcccccc}
%   \hspace{-0.2cm}\\
%   \hline
%               & Acc   & Sens  & Spec  & F1    & MCC   & RMSE  \\ 
%               \cmidrule{2-7}
% Story Training & 0.850 & 0.814 & 0.877 & 0.823 & 0.693 & 1.835 \\
% Story Test     & 0.775 & 0.591 & 0.920 & 0.670 & 0.550 & 2.027 \\
% \hline  
% \end{tabular}
% \caption{\benni{These needs to be extended for the four cases. Case Study. Accuracy, sensitivity, specificity, F1 score, Matthew correlation coefficient (MCC) for gesture phase classification, and residual mean squared error (RMSE) of the observations with respect to the fitted model. \benni{double check RMSE and include log marginal likelihodo}}}
% \label{tab:casestudy_classification_states}
% \end{table}

% Please add the following required packages to your document preamble:
% \usepackage{multirow}

\begin{figure}[htbp]
\centering
\includegraphics[width=1\linewidth]{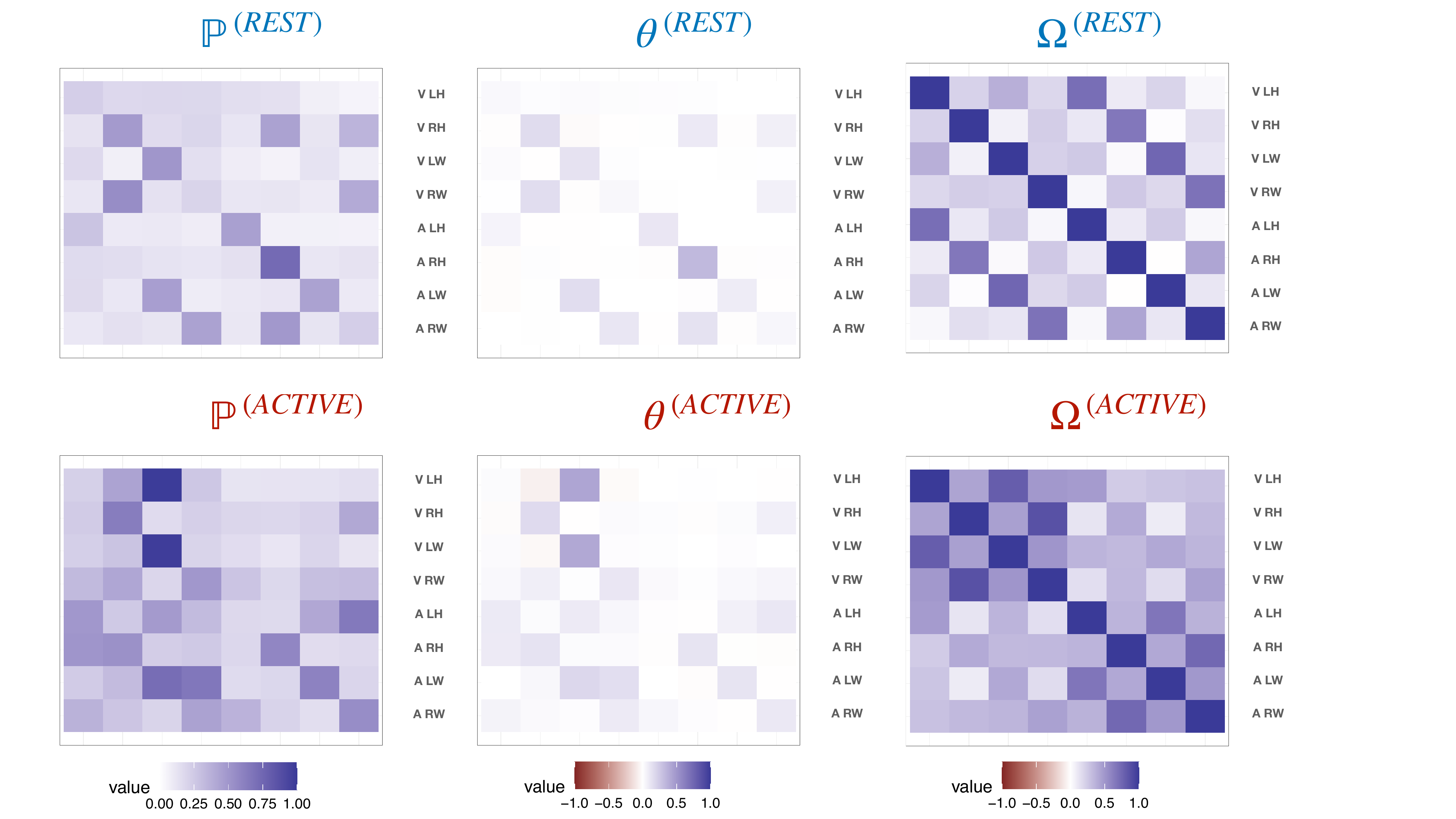}
\caption{Gesture phase application, training story.  Probability of inclusion of the \VAR coefficients (left), their corresponding
magnitude (center), and the correlation matrix (right), for both rest and active state.}
\label{fig:training_summary}
\end{figure}

\begin{figure}[htbp]
\centering
\includegraphics[width=0.7\linewidth]{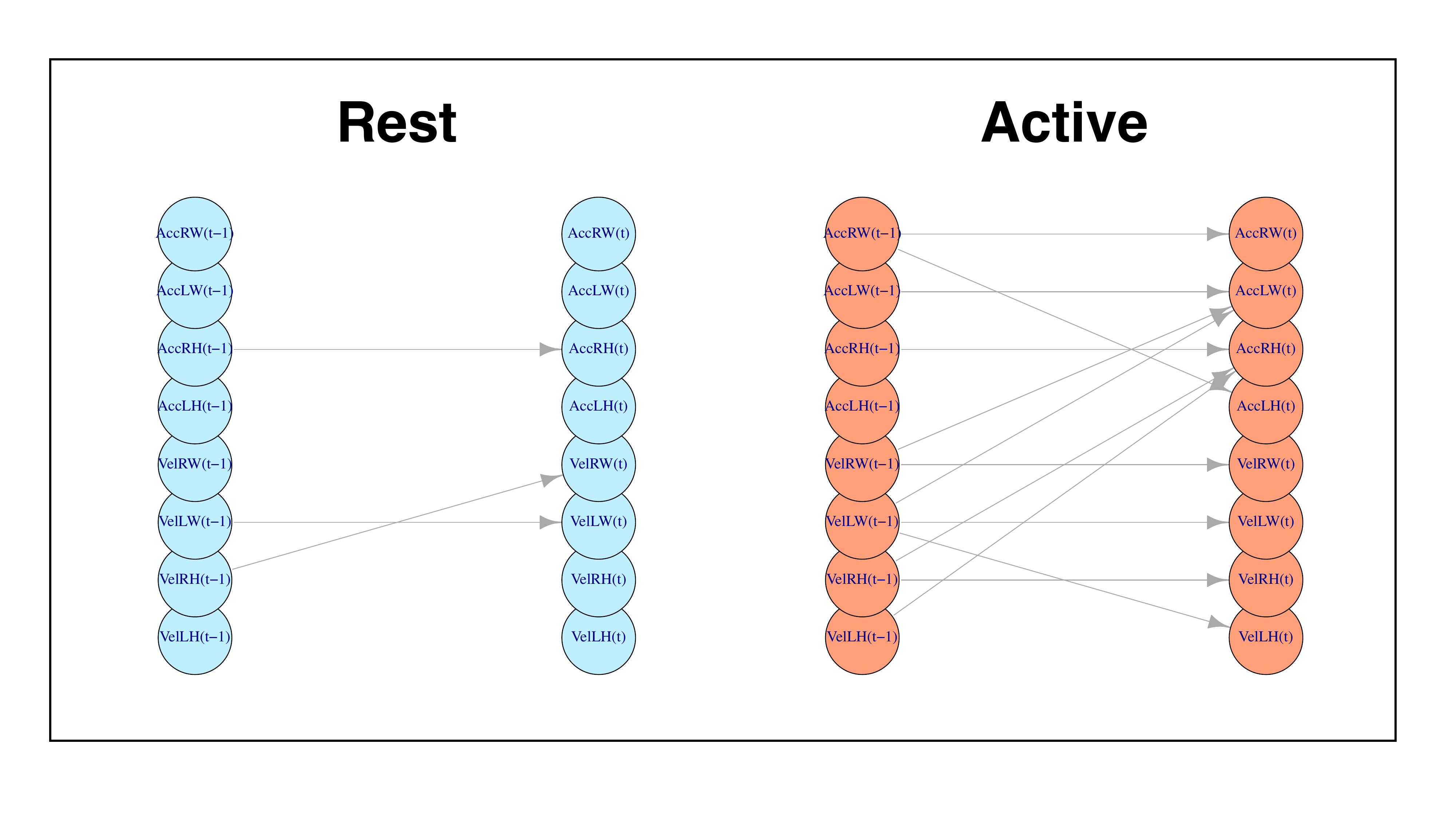}
\caption{Gesture phase application, training story. Active edges of the \VAR matrices, for both rest and active state where the directed edge is drawn if the corresponding probability of inclusion is equal or larger than 0.5.}
\label{fig:training_connection_graph}
\end{figure}

For the proposed model, Figure \ref{fig:training_summary} displays the posterior probability of inclusion of the \VAR coefficients (left), their estimated magnitude (center), and the correlation matrix (right), for both rest and active states (in the training story data).  \textcolor{black}{The temporal nature of \VAR models allows us to investigate the dynamic associations of acceleration and velocity measurements collected from LF, RH, LW, and RW. This task is carried out within each state by quantifying the probability that the associated \VAR parameters are non-zero.}  Figure \ref{fig:training_connection_graph} shows the active edges of the \VAR matrices, for both rest and active states (of the training story data) where the directed edge is drawn if the corresponding probability of inclusion is equal or larger than 0.5 \citep{barbieri2004optimal}. By investigating these plots, we can observe that the activations of the \VAR parameters are different when the individual is actively telling a story and when they are resting. The magnitudes of the \VAR parameters for the resting state appear to be negligible for many entries, as illustrated by the sparsity of the corresponding matrix of probabilities of inclusion; indeed, there are only three edges detected in the corresponding \VAR matrix. Several interesting features can be observed from the matrix of \VAR parameters for the active state: the coefficients relative to the velocities and accelerations, except for LH, are estimated as non-zero (diagonal elements of probabilities of inclusion and \VAR matrices), revealing that accelerations and velocities at time $t-1$ are relevant in characterizing the stream of gestures at time $t$. Furthermore, there are only two nodes with more than one (off-diagonal) edge: for the wrists, acceleration of LW is influenced by the acceleration of RW and LW at time $t-1$;, the velocities of RH and LH at time $t-1$ affect the acceleration of RH at time $t$. It is also noteworthy that the velocity of the RW at time $t-1$ has an edge with a negative magnitude with the velocity of LH at time $t$, suggesting that if, for example, the velocity of RW is high, this will contribute to decreasing the velocity of LH. 

\begin{figure}[htbp]
\centering
\includegraphics[width=0.65\linewidth]{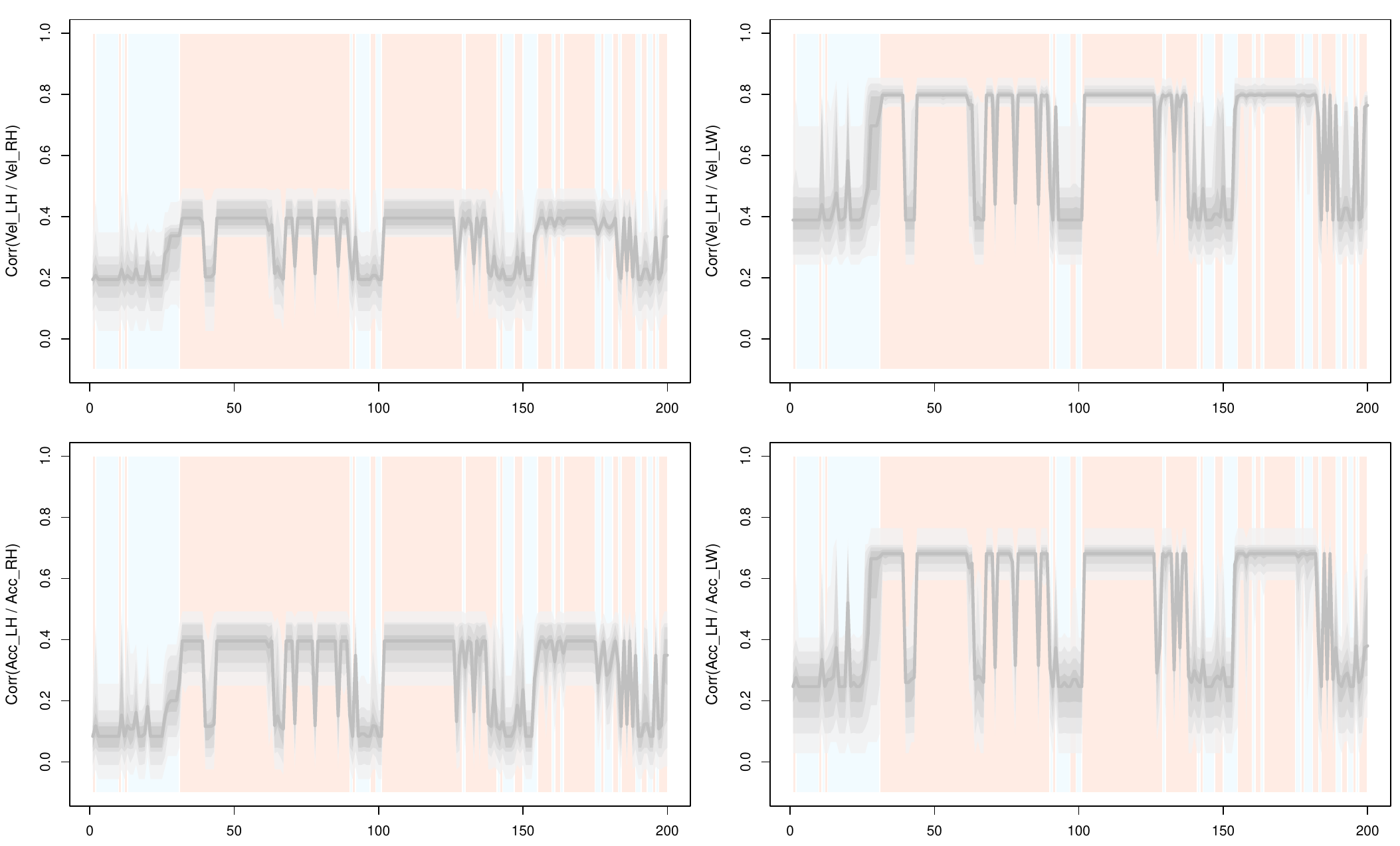}
\caption{Gesture phase application, training story. Posterior predictive time-varying correlation among pairs of measurements, velocity LH and RH, velocity LH and LW, acceleration LH and RH,   acceleration LH and LW. Blue and red vertical bands denote
the estimated resting and active states, respectively. }
\label{fig:time_varying_correlation}
\end{figure}

While the \VAR coefficients capture temporal dependencies, the correlation matrices depict contemporaneous dependencies conditional on the previous time points. The posterior mean estimates for these also show some interesting structure. In the resting state the acceleration and velocity for each location are strongly positively dependent while in the active state there is some block diagonal structure with the blocks corresponding to velocity and acceleration. In particular, greater dependence is estimated between the hand and the wrist on each side. Such findings justify our decision to estimate the full covariance matrix rather than adopt the pseudo likelihood approach of only estimating a diagonal covariance matrix often considered for \VAR models \citep{ghosh2021strong}. We further illustrate  the time-varying behaviour among pairs of measurements (i.e. velocity LH and RH; velocity LH and LW; acceleration LH and RH;   acceleration LH and LW) through posterior predictive  quantiles of their corresponding time-varying correlations, as shown in Figure \ref{fig:time_varying_correlation}. These quantities are obtained by first drawing a state sequence and then conditioned on the drawn sequence, the time-varying sample of the parameter is stored for each \MCMC iteration, so that posterior quantiles can be constructed. The correlations between velocities, and accelerations, of LH and LW (Figure \ref{fig:time_varying_correlation}, top and bottom left) seem to alternate between being slightly correlated for the resting states and being significantly correlated in the active states. Differently, correlations between velocities, and accelerations, of LH and RH (Figure \ref{fig:time_varying_correlation}, top and bottom right) appear to alternate between being slightly positive correlated and mildly correlated, showing that the contemporaneous relationship between left and right hand is not highly marked high, even during gesture activity.

\section{Summary and Discussion} \label{sec:summary_discussion}

In this paper we have proposed a novel vector autoregressive (\VAR)  hidden semi-Markov model (\HSMM) that allows for the modeling of complex dependencies between multivariate time series while enabling the formulation of hidden states associated with different \VAR structures. The explicit-duration semi-Markovian dynamics are represented using a special structure of the transition probability matrix that can approximate the underlying \HSMM with arbitrary accuracy. We have modeled contemporaneous dependencies by estimating a full covariance matrix. Furthermore, we have deployed the recently proposed $l_1$-ball prior to impose sparsity on the state-specific \VAR coefficients and have used a non-local prior construction to conduct selection between the more flexible \HSMM model and the simpler \HMM.
Our simulation results show that our model can well distinguish between different \VAR regimes and that the $l_1$-ball prior provides a powerful and convenient means to characterise sparse structures.
%{\color{red}{One sentence on simulations.} \jack{write something. discussion l1ball does very well in  simulation when it is actually sparse blabla.}}

We have illustrated our proposed methodology in an application to human gesture phase segmentation based on sensor data. Overall, our approach successfully identified periods of rest and active gesturing in a probabilistic way, while also characterizing the existence of switching and temporal dependencies among acceleration and velocity measurements. 
%Under our proposed model, changes in physical activity are expressed through switching of the states as well as via autoregressive patterns characterizing each different state. 
We have further showed that our approach is capable of successfully characterizing and performing predictions for a new video, corresponding to a different, unseen, story.  
%{\color{red}{MV: Talking points from DR (for Intro/Discussion): (1) This prior is computationally attractive as it is amenable to HMC. However, HMC does not provide full posterior inference. (2) This prior reduces to a ``specific" type of spike-and-slab (independent?). The class of SS priors can include more general/flexible constructions.}}

Future work may extend our proposed approach in the following directions. First, the negative-binomial dwell distribution used in this work only allows for the modelling of overdispersed durations (i.e. variance larger than the mean), but there may be scenarios when the state durations show underdispersion. A more flexible family of dwell distribution could be provided by the COM-Poisson \citep{conway1962queuing, benson2021bayesian} distribution, which can account for underdispersed, overdispersed, and equidispersed dwell durations, or the discrete Weibull distribution which ``parsimoniously'' captures over an under dispersion \citep{vinciotti2022bayesian}. However, we note that increased flexibility comes with challenges, since the COM-Poisson prior has an intractable normalizing constant that increases the computational burden. Furthermore, the discrete Weibull only has its first moment in closed form complicating prior elicitation and interpretation. 
Second, it could be of interest to further assume a different degree of sparsity for distinct lags (in addition to  different states) since the magnitude of the \VAR coefficients may decrease as the lag increases. Finally,  our proposed approach does not currently allow for the sparsity of the state-specific covariance matrices. 
%Instead, we could assume the data-generating process to be characterized by a sparse undirected graphical model, for each state. These networks may have many of their nodes not connected to each other, corresponding to entries of the precision matrices that are equal to zero. 
The  $l_1$-ball projection prior may be extended to model state-specific precisions, noting that the sparsity of a precision matrix characterizes the conditional independence among variables.  

\section*{Acknowledgements}

The authors would like to thank David Rossell and Leo Duan for discussions about the $l_1$-ball. B.H-A. and M.V. were supported by NIH-NIMH (R01-MH124115),  J.J. was funded by Juan de la Cierva Formación fellowship  FJC2020-046348-I.

% \benni{Our proposed approach is able to identify both the dynamic properties and stationary structure of multiple signals by simultaneously identifying the active components of the switching VAR matrices and capturing contemporaneous spatial dependencies, while flexibly modelling the general hidden state dwell distribution. }

%\section*{Supplementary Material} 
%Supplementary materials are available and include further details about dwell durations, forecasting functions, graphs of normal pseudo-residuals, and further analysis of the \PA time series results. Code that implements the methodology is available as online supplemental material (see also \url{https://github.com/Beniamino92/BayesianApproxHSMM}).%{https://github.com/Beniamino92/BayesianApproxHSMM})

% \begin{supplement}
% % \sname{Supplement A}\label{suppA} 
% % \stitle{\benni{Title of the Supplement A}}

% We provide supplemental material to the manuscript. \slink[url]{https://github.com/Beniamino92/BayesianApproxHSMM}

% \end{supplement}

%\bibliographystyle{ba}
%\bibliography{bib/biblio}
\bibliography{biblio}

% \section*{Acknowledgements} 

% Beniamino Hadj-Amar and Marina Vannucci would like to thank Jack Jewson to make ISBA a remarkably awful experience. His association has been totally detrimental to their carrier, personal life, and general well-being. Beniamino Hadj-Amar would also like to thank Jack Jewson for ruining his Canada holiday. 

%\end{document}

% \begin{supplement}
% % \sname{Supplement A}\label{suppA} 
% % \stitle{\benni{Title of the Supplement A}}

% We provide supplemental material to the manuscript. \slink[url]{https://github.com/Beniamino92/BayesianApproxHSMM}

% \end{supplement}
\newpage

\appendix

\section{Supplementary Material} 

We provide supplementary material to the manuscript. 
Section \ref{Sec:StateTransitionMatrix} includes the formulation of the special state transition matrix utilized to approximate the \HSMM, as well as an illustrative example of such a construction. In Section \ref{Sec:l1ball_projection_alg} we provide the details for the $l_1$-ball projection algorithm.  Section \ref{Sec:Add_Sims} contains additional results of the simulation studies, and we illustrate further results of the gesture phase application in Section \ref{Sec:AdditionalGesturePhase}.

\subsection{State Transition Matrix}{\label{Sec:StateTransitionMatrix}}

The matrix $\bm{A}$, introduced in Section \ref{seq:switching_model}, has the following structure
\begin{equation}
\label{eq:phi_mat}
\bm{A} = \begin{bmatrix} 
\bm{A}_{11} & \dots & \bm{A}_{1K} \\
\vdots & \ddots & \vdots \\
\bm{A}_{K1} & \dots       & \bm{A}_{KK} 
\end{bmatrix},
\end{equation} \begin{equation}
\bm{A}_{jj} = 
\begin{bmatrix}
0 & 1 - \phi_j\,(1) & 0 & \dots & 0 \\
\vdots & 0 & \ddots &  & \vdots \\
&\vdots &  &  & 0 \\
0 & 0 & \dots & 0 & 1 - \phi_j\,(b_j - 1) \\
0 & 0 & \dots & 0 &  1 - \phi_j\,(b_j)
\end{bmatrix}, \quad 
\bm{A}_{jk} = 
\begin{bmatrix}
\pi_{jk}\, \phi_j\,(1) & 0 & \dots & 0 \\
\pi_{jk}\, \phi_j\,(2) & 0 & \dots & 0 \\
\vdots &  &  & \\
\pi_{jk} \, \phi_j\,(b_j) & 0 & \dots & 0
\end{bmatrix},
\end{equation}
where the sub-matrices $\bm{A}_{jj}$ along the main diagonal, of dimension $b_j \times b_j $, are defined for $b_j \geq 2$, and $\bm{A}_{jj} = 1 - \phi_j(1)$, for $b_j = 1$. The $b_j \times b_k$ off-diagonal matrices $\bm{A}_{jk}$ are 0 expect for the first column which contains \textit{hazard rates} defined for $ r \in \mathbb{N}_{> 0} $
\begin{equation}
\phi_j \, (r) = \dfrac{p \, (\, d_j = r \, | \,  \bm{\lambda}_j)}{p \, (\, d_j \geq r \, | \,  \bm{\lambda}_j)}, \quad \text{if} \, \, p \, (\, d_j \geq r -1  \, | \,  \bm{\lambda}_j) < 1,
\label{eq:hazard_rates}
\end{equation}
and 1 otherwise, where $p \, (\, d_j = r \, | \,  \bm{\lambda}_j)$  denotes the probability mass function of the dwell distribution $g \, ( \bm{\lambda}_j)$ for state $j$. 
This structure for the matrix $\bm{A}$ implies that transitions within state aggregate $B_j$ are determined by diagonal matrices $\bm{A}_{jj}$, while transitions between state aggregates $B_j$ and $B_k$ are controlled by off-diagonal matrices $\bm{A}_{jk}$. 

% Figure \ref{fig:HSMMapprox_vs_HMM_example} illustrates an example of formulating an \HSMM through an \HMM, where for the sake of the illustration, this representation is made exact, namely there is no approximation. A two-state \HSMM characterized by emission distributions $f(\bm{\theta}_j)$, and dwell-duration distributions $d_j(r)$, for  $j=1, 2$, is represented as an \HMM with extended state space $\bar{B} = 7$ with dwell-threshold $\bm{b} = (4, 3)$ for state aggregates $B_1$ and $B_2$, respectively. Note that the dwell-thresholds are selected to cover the entire support of the duration distributions, so that this \HMM formulation corresponds to its exact \HSMM representation. The transition matrix $\bm{A}$ is  completely determined by the \HSMM dwell distribution $p \, (\, d_j = r )$ (within state aggregate moves) and the between state aggregate transition probabilities $\pi_{12}$ and $\pi_{21}$, which in this case are equal to one, since we consider a two-state \HSMM.   

%  \begin{figure}[htbp]
%  \centering
% \includegraphics[width=0.75\linewidth]{figures/HSMMapprox_vs_HMM.pdf}
% \caption{Example of formulating an \HSMM through an \HMM,}
% \label{fig:HSMMapprox_vs_HMM_example}
% \end{figure}

\subsection{$l_1$-Ball Projection Algorithm} \label{Sec:l1ball_projection_alg}

The algorithm of \cite{xu2020bayesian} to compute the $l_1$-ball projection solving \eqref{eq:l1_ball_2} is provided in Algorithm \ref{Alg:l1_ball}

\begin{algorithm}
\caption{$l_1$-ball projection. $\, \, $ \textit{\textbf{Input:}}
$\bm{\beta}_p^{\,j} \in \mathbb{R}^{D^2}$ and $r^{\,j} \in \mathbb{R}_{+}$. $\, \, $ \textit{\textbf{Output:}}
$\bm{\Theta}_p^{\,j} \in \mathbb{R}^{D \times D} \cup \{0\}$ }\label{alg:l1_balla}
\begin{algorithmic}
\If{$||\bm{\beta}_p^{\,j}||_1 \leq r^{\,j}$}
    \State $\bm{\Theta}_p^{\,j} \gets vec^{-1}(\bm{\beta}_p^{\,j})$
\Else    
\State Sort $\bm{\beta}_p^{\,j}$ so that $|\beta^{\,j}_{p\,(1)}| \geq \cdots \geq |\beta^{\, j}_{p\,(D^2)}|$ 
\State $\phi_l\gets \left(\sum_{n=1}^l |\beta^{\,j}_{p\,(n)}| - r^{\,j}\right)_{+},$ $\, \qquad \quad \text{for } \, l = 1, \dots, D^2. $ 
\State $m \gets \max\left\{n: |\beta^{\,j}_{p\,(n)}| > \frac{\phi_n}{n}\right\}$
\State $\tilde{\phi} \gets \frac{\phi_m}{m}$
\State $\theta^{\,j}_{p i} \gets~ \textup{sign}(\beta^{\,j}_{pi})\, \max\left(|\beta^{\,j}_{pi}| - \tilde{\phi}, 0 \right),$ $\, \, \text{for } \, l = 1, \dots, D^2. $ 
\State $\bm{\Theta}_p^{\,j} \gets vec^{-1}(\bm{\theta}_p^{\,j})$
\EndIf
\end{algorithmic}
\label{Alg:l1_ball}
\end{algorithm}

While \stan{} and \NUTS{} attempt to adapt sampling to the geometry of the posterior, reparametrizations of the sample space can help to facilitate this in high dimensions: i) we reparametrized the latent variables for the \VAR coefficients as $\tilde{\bm{\beta}}^{\, j} := \frac{\bm{\beta}^{\, j}}{\sqrt{T}}$ motivated by the fact that as the sample size grows the posterior of the $\bm{\beta}^{\, j}$ does not concentrate% (see Section \ref{Equ:limiting_l1})
;  ii) we reparameterized the $l_1$-ball radius as $\tilde{r}^{\, j} := \frac{r^{\, j}}{\sqrt{D^2P}}$ motivated by the fact that the radius is learned from the $D^2P$ elements of $\bm{\theta}^{\, j}$; iii) we found that reparametrizing dwell parameters $\tilde{p}^{\, j} := \frac{\lambda^{\, j}}{1 + \lambda^{\, j}}$ and $\tilde{q}^{\, j} := \frac{\rho^{\, j}}{1 + \rho^{\, j}}$ for the \HSMM, more closely resembling the parameterization of the \HMM, also improved performance. The above reparametrizations facilitate posterior sampling while leaving the model unchanged.

\subsection{Additional Simulation Results}{\label{Sec:Add_Sims}}

\begin{figure}[ht!]
\centering
\includegraphics[width=0.32\linewidth]{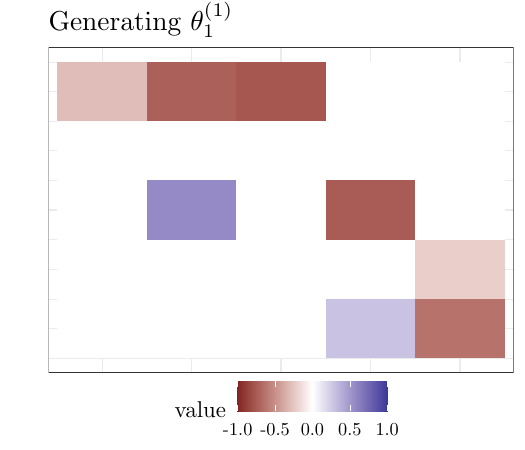}
\includegraphics[width=0.32\linewidth]{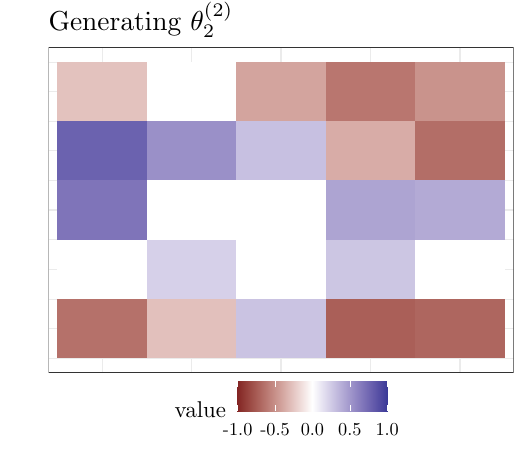}
\includegraphics[width=0.32\linewidth]{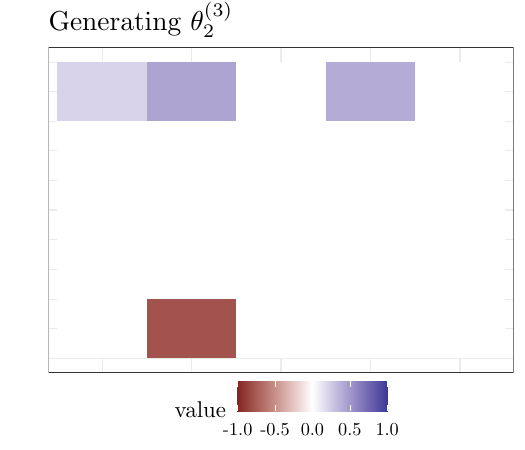}\\
\includegraphics[width=0.32\linewidth]{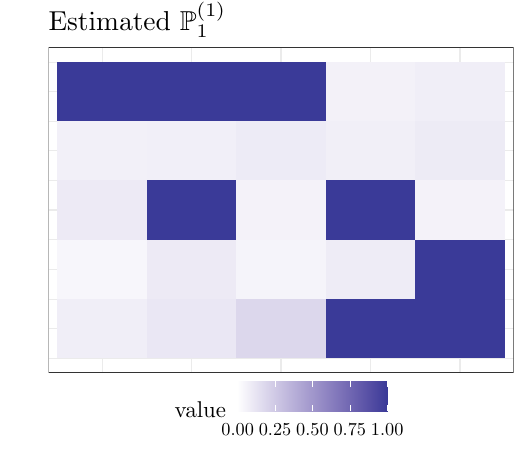}
\includegraphics[width=0.32\linewidth]{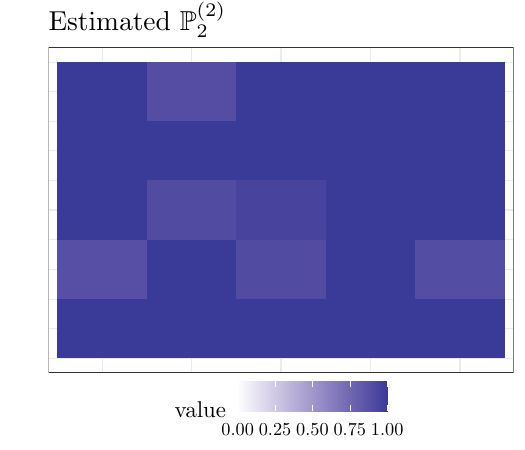}
\includegraphics[width=0.32\linewidth]{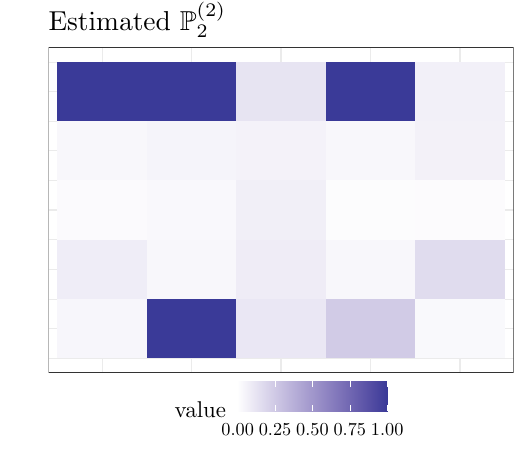}\\
\caption{Simulation Study. Generating \VAR parameters (order 2) and their estimated posterior probability of inclusion.}
\label{fig:Simulation_VAR_inclusion_order2}
\end{figure}

For the simulated data as described in Section \ref{Sec:HSMM_sims}, Figure \ref{fig:Simulation_VAR_inclusion_order2} compares the estimated inclusion probabilities of the order 2 \VAR coefficients in each regime, $\bm{\Theta}_2^{\, j}$, to their generating values. Similarly to the order 1 estimates, all non-zero generating parameters are estimated to have close to 1 probability of inclusion and most truly 0 estimates have a small probability of inclusion. The exception is still State 2 where the generating \VAR estimates were not particularly sparse.

%simulation_VARHMM_HSMM_L1ball.Rmd

\begin{figure}[ht!]
\centering
\includegraphics[width=0.32\linewidth]{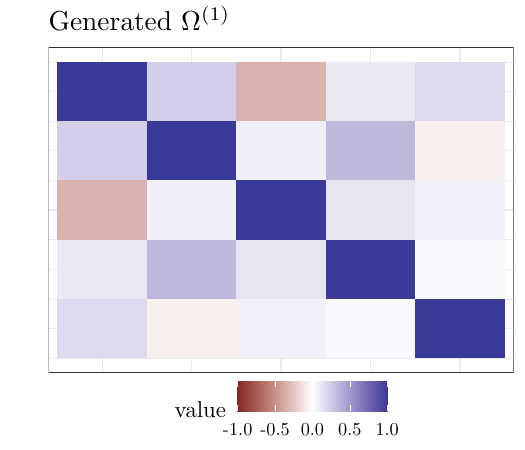}
\includegraphics[width=0.32\linewidth]{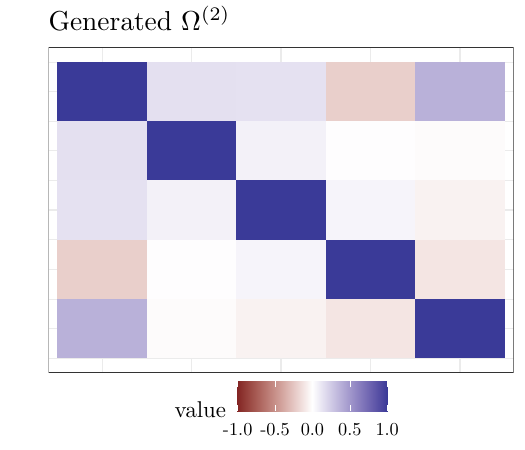}
\includegraphics[width=0.32\linewidth]{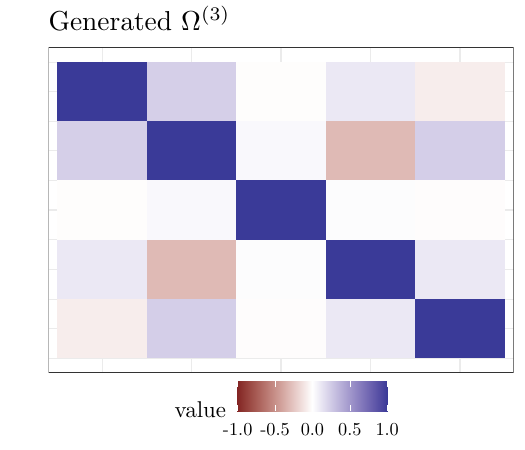}\\
\includegraphics[width=0.32\linewidth]{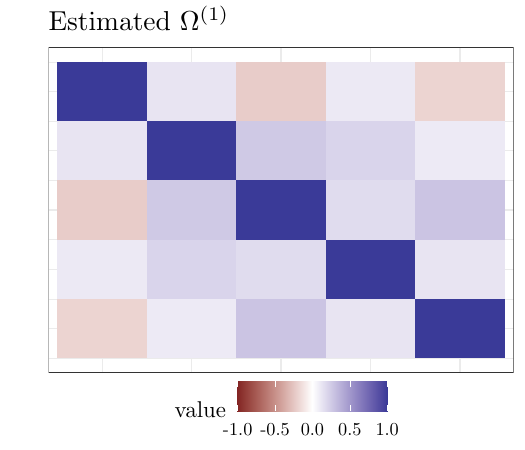}
\includegraphics[width=0.32\linewidth]{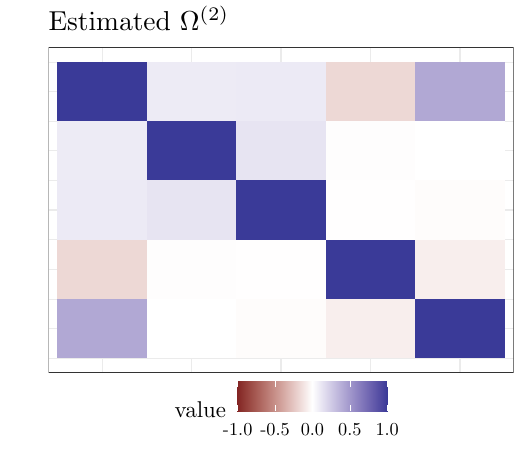}
\includegraphics[width=0.32\linewidth]{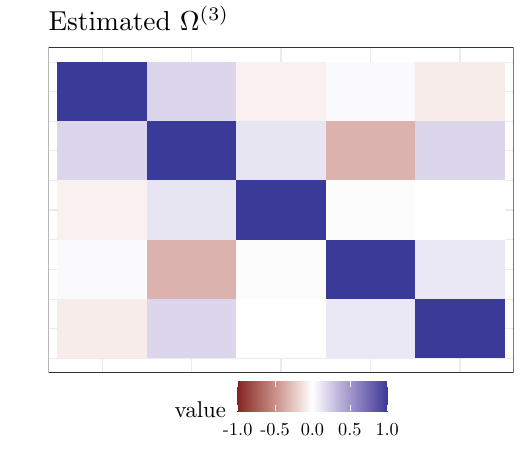}\\
\caption{Simulation Study. Generating correlation matrices and their estimated posterior means.}
\label{fig:Simulation_Omega}
\end{figure}

Figure \ref{fig:Simulation_Omega} compares the posterior mean estimates for $\bm{\Omega}^{\, j}$ with the data generating values. As the data was standardised, we compare estimated and generated values for pairwise correlations $\bm{\Omega}^{\, j}$ rather than covariances $\bm{\Sigma}^{\, j}$. For States 2 and 3 the posterior mean estimate for correlation matrix $\bm{\Omega}^{\, j}$ very closely resemble the generating values while the estimates for State 1 are reasonable. 

\begin{figure}[!ht]%[ht!]
\centering
\includegraphics[width=0.49\linewidth]{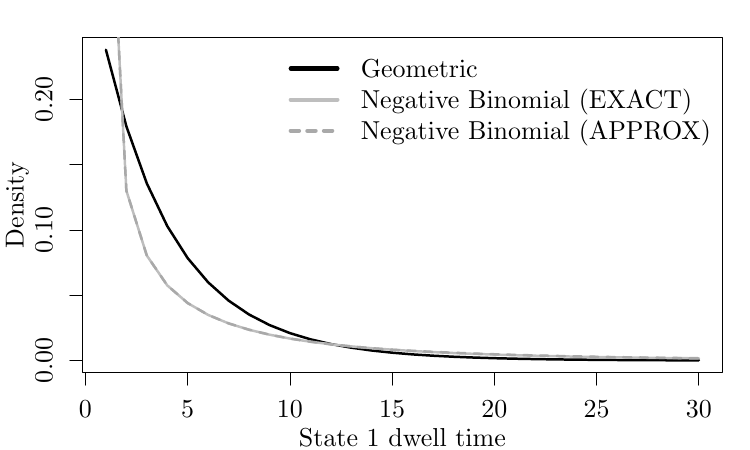}
\includegraphics[width=0.49\linewidth]{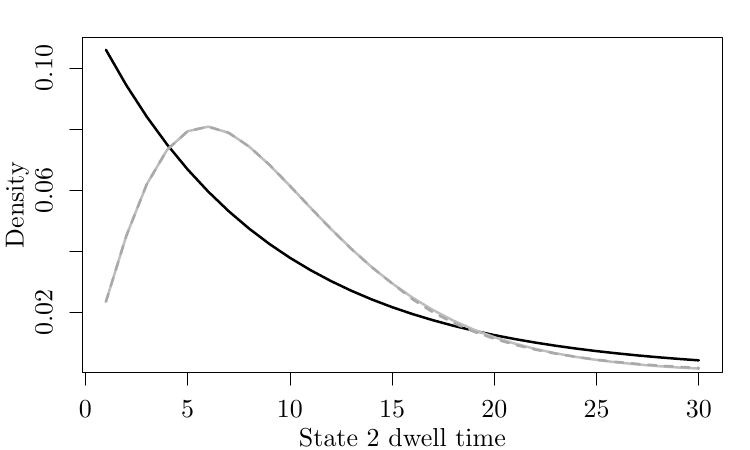}\\
\includegraphics[width=0.49\linewidth]{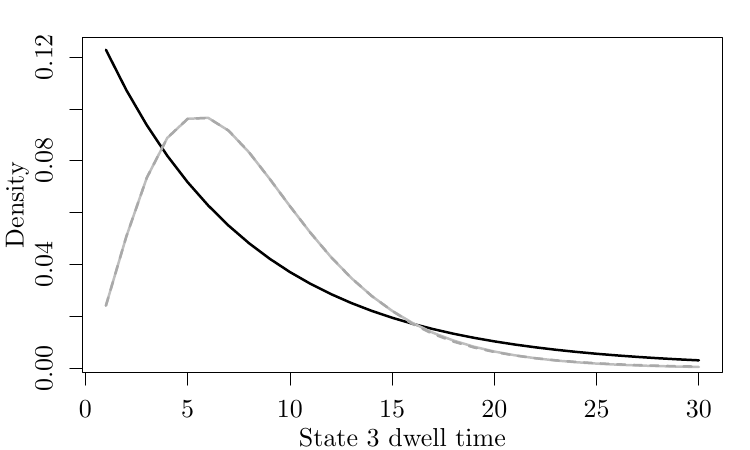}
\caption{Simulation Study. Posterior predictive dwell distributions for each states under the \HMM (geometric dwell), the approximation to the negative binomial \HSMM resulting from $\bm{b} = (15, 15, 15)$ and the exact negative binomial \HSMM dwell distribution corresponding to the same parameter posterior.}
\label{fig:Simulation_Dwell}
\end{figure}

Finally, Figure \ref{fig:Simulation_Dwell} plots the posterior predictive distributions for the dwell distribution in each state estimated under the \HMM (geometric dwell) and the negative binomial \HSMM. For the \HSMM we plot both the posterior predictive for our model's approximation to the \HSMM dwell and the exact negative binomial dwell that would correspond to the same parameters. Such a plot demonstrates that taking $\bm{b} = (15, 15, 15)$ was sufficiently large to provide negligible approximation of the fitted dwell distribution.

%While Bayesian model selection is known to consistently select the data generating model, the rate at which is selects the simpler of two nested models can be considerably slower than the rate at which it selects the more complicated one. This is particularly worrisome when estimating the dwell distribution in \HSMMs as the number of state transitions in a time series is often many fewer than the number of observations. 
%To illustrate this we conducted a second simulation exercise,

For completeness, we conducted a further simulation exercise, this time simulating a $D = 5$ dimensional time series of $T = 200$ from a $K = 2$ state \HMM with \VAR emission distribution of order $P = 1$ in each regime. The self transition probabilities in each state were set to $\tilde{\pi}_{jj} = 0.9$ for $j = 1, 2$. The \VAR parameters, intercepts and covariance in each regime were generated according to the procedure outlined as in the previous simulation, where the \VAR had 70\% sparsity in State 1 and 30\% sparsity in State 2. We repeated the simulation of the time series 10 times. For each simulation we considered selecting between the generating \HMM model and the more complex \HSMM models under the local and non-local priors. %While Bayesian model selection is known to consistently select the data generating model, the rate at which is selects the simpler of two nested models can be considerably slower than the rate at which it selects the more complicated one

%simulation_VARHMM_HSMM_L1ball.Rmd
\begin{figure}[ht!]
\centering
\includegraphics[trim= {0.0cm 0.5cm 0.0cm 0.0cm}, clip,width=0.8\linewidth]{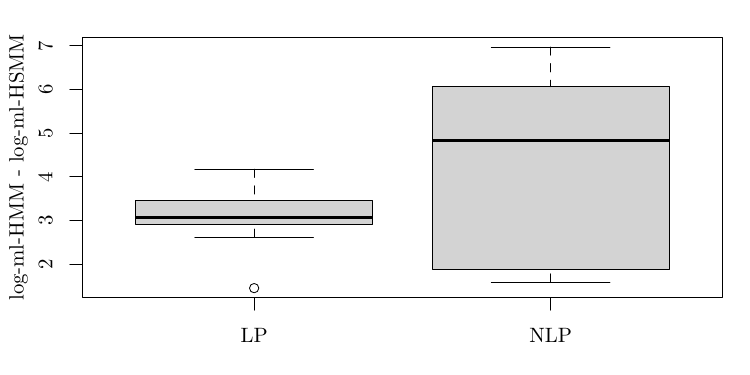}
\caption{Simulation Study. Comparison of \HMM-$l_1$ log-ml with  \HSMM-$l_1$ and \HSMM-$l_1$-\NLP across 10 repeat datasets drawn from the simpler \HMM model. Positive values indicate correctly selecting the simpler \HMM model.}
%trim={<left> <lower> <right> <upper>}
\label{fig:HMM_logml}
\end{figure}

Figure \ref{fig:HMM_logml} compares the difference in log-ml of the \HMM-$l_1$ model with that of the \HSMM-$l_1$ and \HSMM-$l_1$-\NLP models across the 10 repeats. Under the \NLP the difference in log-ml between the \HMM and \HSMM is more positive, indicating that the model selection decision correctly favours the simpler \HMM model to a greater extend. However, under the \LP this difference is also always positive indicating that the correct model selection decision is still made here. This is particularly reassuring as the number of state transitions in a time series is often many fewer than the number of observations and so \HMM/\HSMM dwell distribution are often estimated using few data points.
%To illustrate this we conducted a second simulation exercise,

\subsection{Additional Gesture Phase Segmentation Results}{\label{Sec:AdditionalGesturePhase}}

\begin{figure}[htbp]
\centering
\includegraphics[trim= {0.0cm 6.00cm 0.0cm 0.0cm}, clip,width=1.0\linewidth]{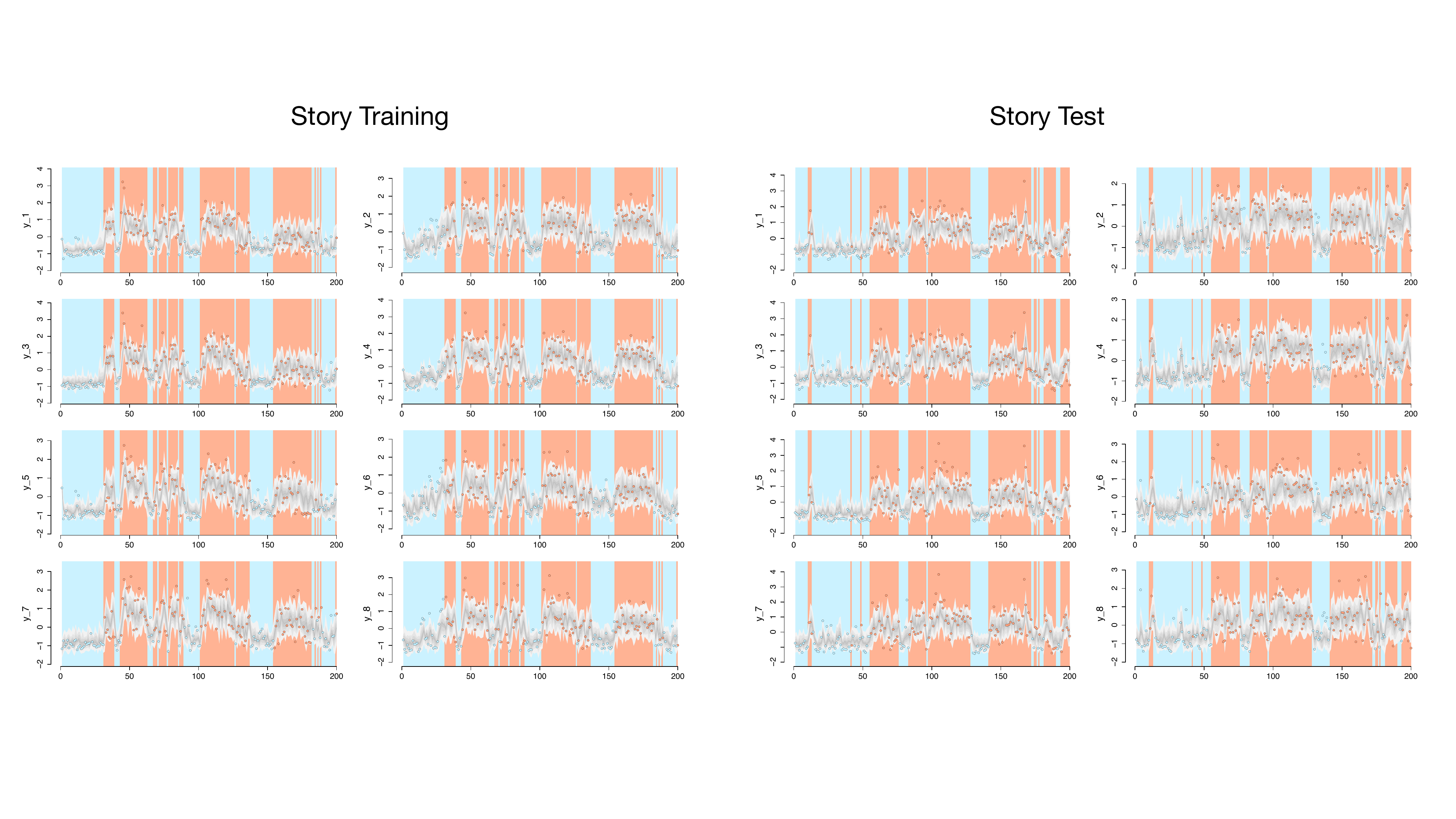}
%trim={<left> <lower> <right> <upper>}
\caption{Gesture phase application. Graphical posterior predictive check consisting of the observations alongside 100 draws from the estimated posterior predictive, for both training story (left) and test story (right). }
\label{Fig:GesturePhase_marginals}
\end{figure}

For the gesture phase segmentation application described in Section \ref{Sec:GesturePhase}, Figure \ref{Fig:GesturePhase_marginals} plots the individual dimensions of the training and testing data, shown jointly in Figure \ref{fig:data_training_vs_test}, along side 100 draws from the estimates posterior predictive and the MAP state classification.

\begin{figure}[ht!]
\centering
\includegraphics[width=0.49\linewidth]{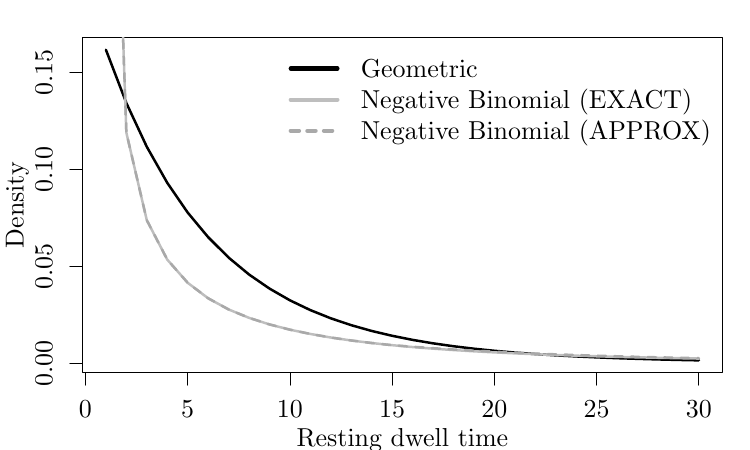}
\includegraphics[width=0.49\linewidth]{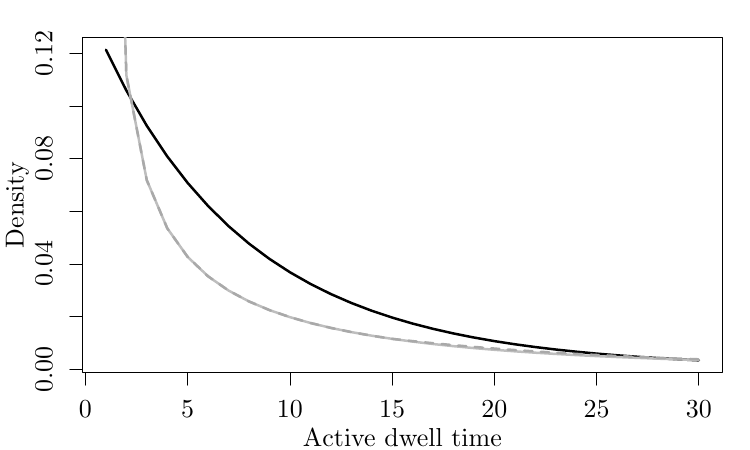}
\caption{Gesture phase application. Posterior predictive dwell distributions for each states under the \HMM (geometric dwell), the approximation to the negative binomial \HSMM resulting from $\bm{b} = (15, 15)$ and the exact negative binomial \HSMM dwell distribution corresponding to the same parameter posterior.}
\label{fig:Simulation_Dwell_Gesture}
\end{figure}

Figure \ref{fig:Simulation_Dwell_Gesture} is analagous to Figure \ref{fig:Simulation_Dwell} and plots the posterior predictive distributions for the dwell distribution in each state estimated under the \HMM (geometric dwell) and the exact and approximate negative binomial \HSMM. This demonstrates that taking $\bm{b} = (15, 15, 15)$ was sufficiently large to provide negligible approximation of the fitted dwell distribution.

\begin{figure}[htbp]
\centering
\includegraphics[width=0.49\linewidth]{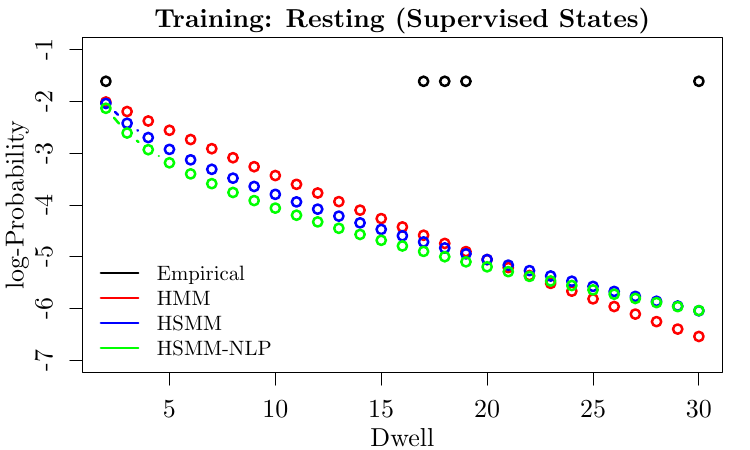}
\includegraphics[width=0.49\linewidth]{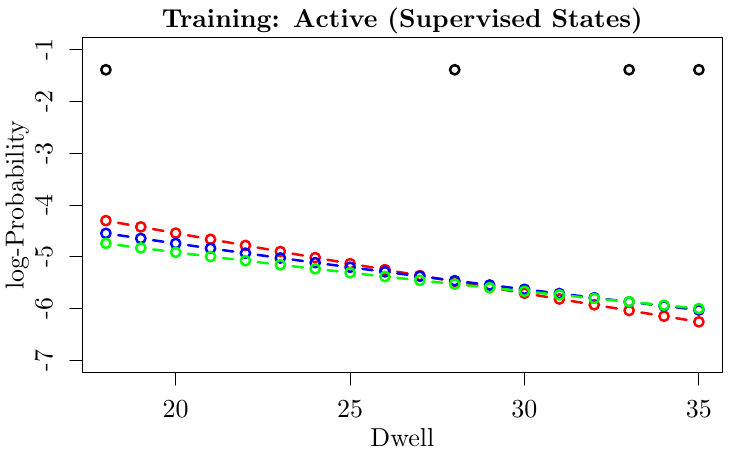}\\
\includegraphics[width=0.49\linewidth]{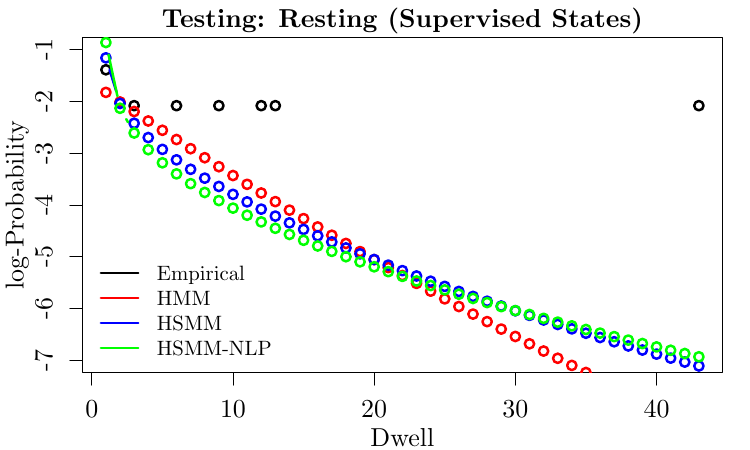}
\includegraphics[width=0.49\linewidth]{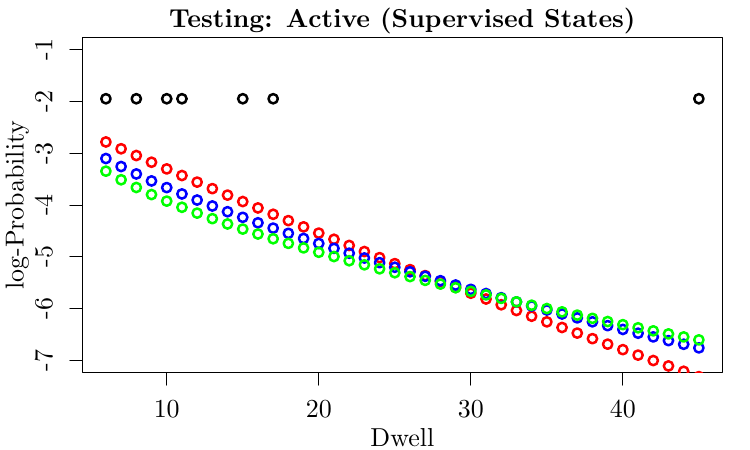}
\caption{Comparison of the estimated dwell distributions under the \HMM, \HSMM and \HSMM-\NLP with the empirical distribution of the supervised states in resting and active states in \textit{training} (\textbf{top}) and \textit{testing} (\textbf{bottom}) on the log-probability scale.}
\label{fig:empirical_dwells}
\end{figure}

\textcolor{black}{
We plotted the estimated dwell distributions under the \HMM, \HSMM and \HSMM-NLP and compared these with the empirical dwell distributions provided by the supervised states for both the training and testing data in Figure \ref{fig:empirical_dwells}. We see here that the \HSMM-\NLP is able to fit a heavier tailed dwell distribution (corresponding to $\rho < 1$) and is able to assign higher mass to the longer dwell times associated with the supervised states. 
}

\begin{figure}[htbp]
\centering
\includegraphics[width=0.49\linewidth]{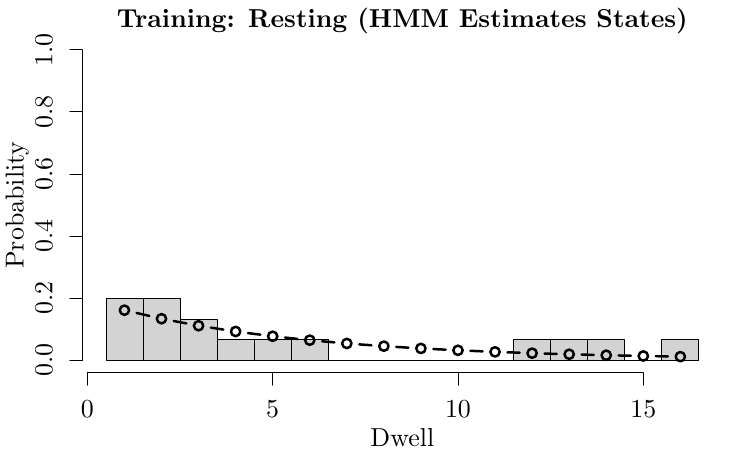}
\includegraphics[width=0.49\linewidth]{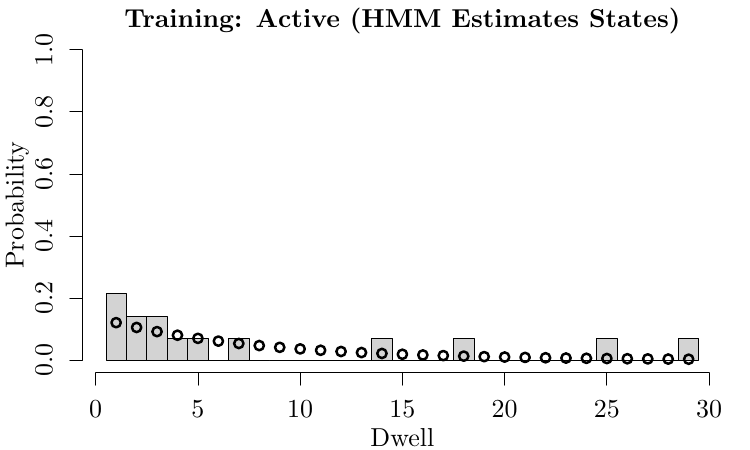}\\
\includegraphics[width=0.49\linewidth]{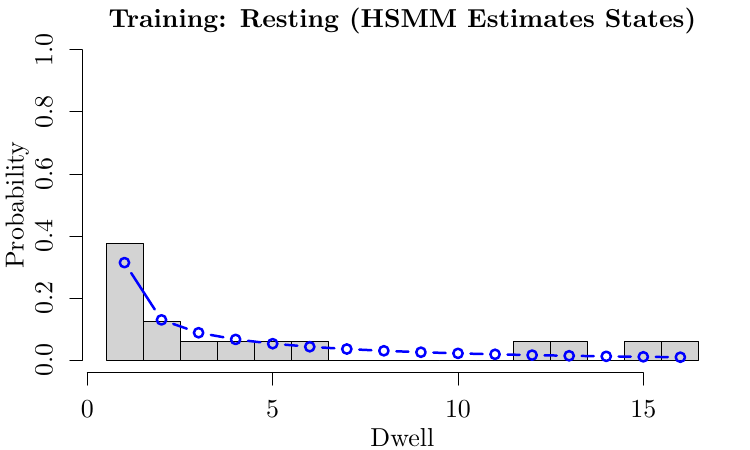}
\includegraphics[width=0.49\linewidth]{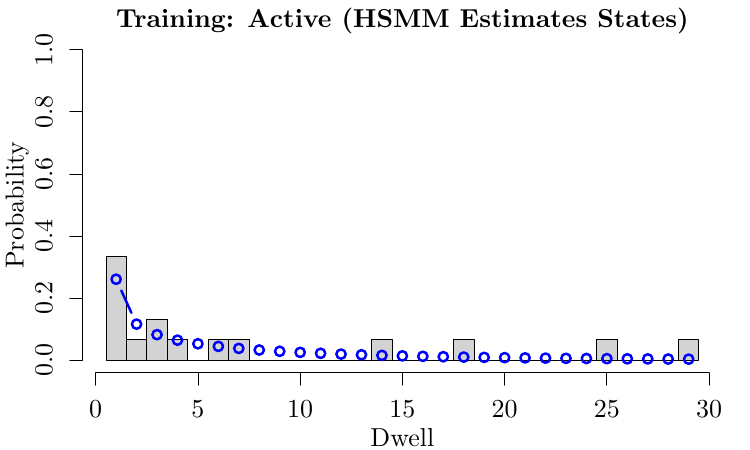}\\
\includegraphics[width=0.49\linewidth]{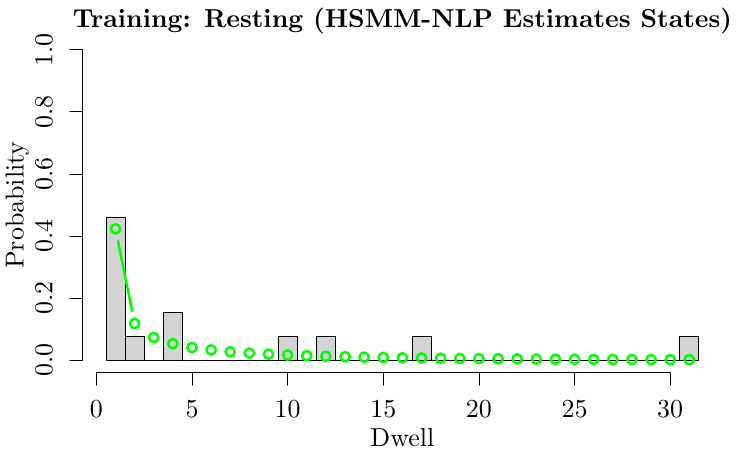}
\includegraphics[width=0.49\linewidth]{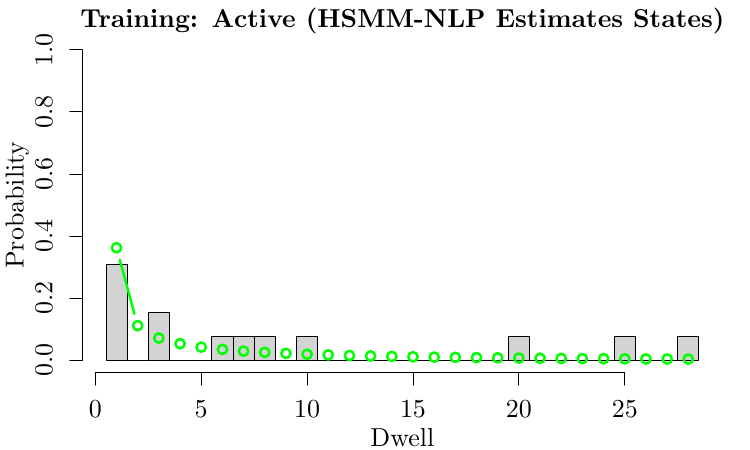}\\
\caption{Comparison of the estimated dwell distributions under the \HMM, \HSMM and \HSMM-\NLP alongside the empirical distribution of the dwell times in the MAP estimated state sequence for the \textit{training} data.}
\label{fig:viterbi_empirical_dwells}
\end{figure}

\begin{figure}[htbp]
\centering
\includegraphics[width=0.49\linewidth]{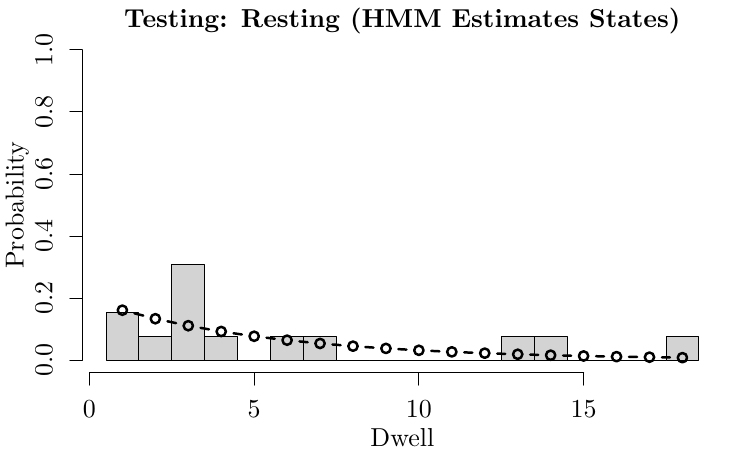}
\includegraphics[width=0.49\linewidth]{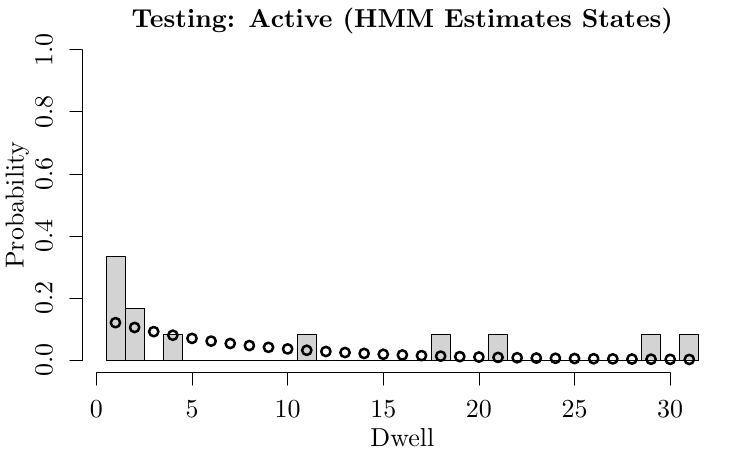}\\
\includegraphics[width=0.49\linewidth]{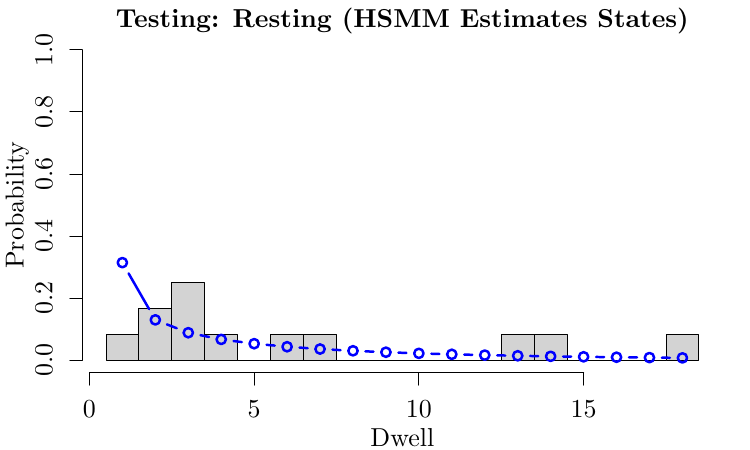}
\includegraphics[width=0.49\linewidth]{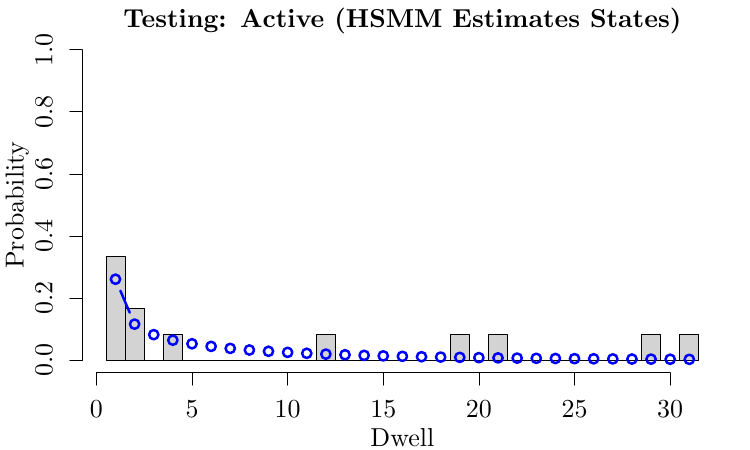}\\
\includegraphics[width=0.49\linewidth]{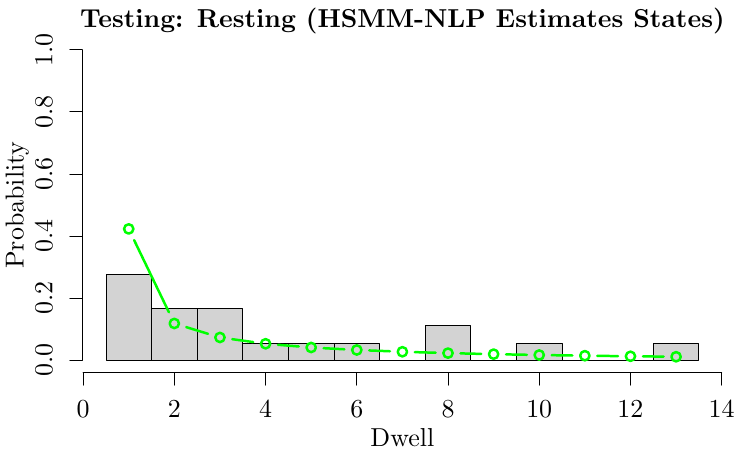}
\includegraphics[width=0.49\linewidth]{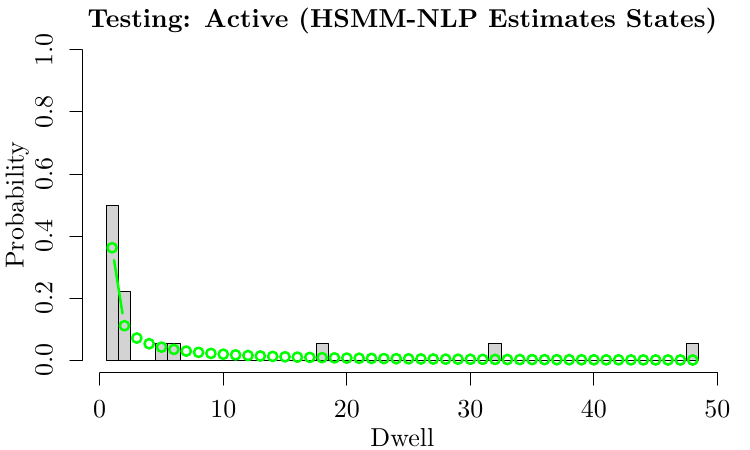}\\
\caption{Comparison of the estimated dwell distributions under the \HMM, \HSMM and \HSMM-\NLP alongside the empirical distribution of the dwell times in the MAP estimated state sequence for the \textit{testing} data.}
\label{fig:viterbi_empirical_dwells_test}
\end{figure}

\textcolor{black}{
Remembering, that our model treats the states as unlabelled, Figures \ref{fig:viterbi_empirical_dwells} and  \ref{fig:viterbi_empirical_dwells_test} further plots the estimated dwell distribution under the \HMM, \HSMM and \HSMM-NLP and compared this to the empirical distributions of the dwells times in their MAP estimated state sequence for the training and testing sets respectively. We see that the increased mass on longer dwells observed in Figure \ref{fig:empirical_dwells} allows the \HSMM-\NLP to estimate longer dwell times than the other methods and achieve better correspondence with the supervised state assessment. 
}

\end{document}